\documentstyle[aps,pre,epsf]{revtex}
\newcommand{\Sq}{{\bf S}(q)} 
\newcommand{\Fq}{{\bf F}(q)} 
\newcommand{\Mq}{{\bf {\cal F}}(q)} 
\input{epsf}

\begin{document} 
\draft 
\title{\bf Molecular mode-coupling theory applied to a liquid of
diatomic molecules}

\author{A. Winkler, A. Latz\footnote{author to whom correspondence
should be addressed}, R. Schilling, C. Theis$^1$}

\address{Institut f\"ur Physik, Johannes
Gutenberg--Universit\"at, Staudinger Weg 7, D-55099 Mainz, Germany \\
$^1$Fakult\"at f\"ur Physik, Albert--Ludwigs--Universit\"at,
Hermann--Herder--Stra{\ss}e 3, D--79104 Freiburg, Germany}

\date{\today} 
\maketitle

\begin{abstract} 
We study the molecular  mode coupling theory for a liquid of diatomic 
molecules.  The equations for the critical 
tensorial nonergodicity parameters ${\bf F}_{ll'}^m(q)$ 
and the critical amplitudes of the $\beta$ -
relaxation  ${\bf H}_{ll'}^m(q)$ are solved up to a cut off $l_{co}$ = 2 
without any further
approximations. Here $l,m$ are indices of spherical harmonics.  
Contrary to previous studies, where additional
approximations were  applied, we find in agreement with simulations,
that all molecular degrees of freedom vitrify
at a single temperature $T_c$.  The theoretical results 
for the non ergodicity parameters and the critical amplitudes are
compared with those from simulations.  
The qualitative agreement is good for all
molecular degrees of freedom. 
To study the influence of the cut
off on the non ergodicity parameter, we also calculate the non ergodicity 
parameters for an upper cut off $l_{co}=4$.  In addition we also
propose a new method for the calculation of the critical nonergodicity 
parameter 
\end{abstract}
%


\section{Introduction} \label{sec:introduction}
The mode coupling theory (MCT) of the glass transition is by now an
important tool to understand experiments in and simulations of supercooled
liquids \cite{foot1}. For a long time most of the theoretical investigations
concentrated on simple monoatomic or binary liquids. All universal
and even system specific predictions of these  investigations could
be tested on a {\it quantitative} level in a system of hard
colloids \cite{colloid1a,colloid1}, which
is an excellent realization of a hard sphere system and in computer
simulations for a binary Lennard Jones system \cite{kob_andersen,kob_nauroth}. 
Details of theory and tests for simple glass formers 
can be found in review articles
\cite{leshouches,repphys,schilling-review,yip,pisa_go,pisa_cu,kob_rev}
and articles
cited therein. 

Although the theory was originally formulated only for these
simple systems, most of the experimental and simulation support came
from research 
on much more complex system (e.g. 
tri-$\alpha$-naphtylbenzene \cite{naphta},  
Orthoterphenyl (OTP) \cite{otp1,otp2,otp3,otp4},
$0.4Ca(NO_3)_20.6KNO_3$ (CKN) \cite{CKN0,CKN1,CKN2,CKN3}, 
Glycerol \cite{glycerol1,glycerol2,glycerol3}, Salol
\cite{salol1,salol2,salol3}, toluene \cite{toluene1} and water
\cite{water1,water2}).  
Also most of the experimental
methods used, did {\it not} measure density correlation functions or
their susceptibilities, for which the original theory was formulated. Even
neutron scattering experiments in systems consisting of molecules
whose components have different cross sections for neutrons
\cite{otp4,glycerol2,toluene1} do not measure the density
correlation function exclusively, but a mixture of more complicated
correlation functions involving molecular degrees of freedom 
\cite{neutrontheory} (see  also \cite{single1} for a single linear
molecule).  Dielectric loss measurements
\cite{CKN3,glycerol3} measure
directly the correlation function of a tensor of rank $1$. Depolarized
light scattering \cite{otp1,CKN1,salol1,cummrot}, Kerr effect experiments
\cite{torre_kerr,hinze_kerr}, NMR \cite{sillescu,rossler,spiess}and
ESR \cite{leporini} (and references therein) measure correlation
function of a tensor
of rank $2$. The mentioned tensorial quantities are all related to 
orientational degrees of freedom (ODOF), whereas the original theory
\cite{bgs} 
only considered translational degrees of freedom (TDOF), i.e. the
center of mass motion.  But of course, when comparing 
experimental results on 
complex systems with predictions of the MCT for  simple liquids it
was always reasonable to argue, that there are in  every experiment
couplings to the center of mass motion. E.g. the reorientation of
dipoles measured in dielectric loss measurements can induce center of
mass motion via a translation - rotation coupling. Also the
reorientation of the polarizability tensor in light scattering
measurements is related to a physical rotation of the molecules and
will therefore be coupled to the center of mass motion of the
molecules as well. A slowing down of this motion due to very slow structural
relaxations can consequently also indirectly be measured in the mentioned
experiments. In addition it is perfectly justified to perform tests of
the {\it universal} predictions of MCT in complicated molecular and
polymeric systems \cite{bennemann} for   
$\beta$ scaling laws and properties of the $\alpha$ relaxation, like
time temperature superposition principle and wave vector dependent
stretching exponents, since the underlying universal features of the
bifurcation scenario should also remain valid for molecular systems.  

But beyond the universal aspects MCT aims to be a microscopic
theory of structural relaxation. This goal  is to a large extent
achieved for simple liquids.  There it was possible to obtain
quantitative agreement between experiment and theory for the full
dynamic range of structural relaxation (i.e. $\beta$ and $\alpha$ -
relaxation) \cite{colloid1a,colloid1,kob_nauroth}. 
Recently also  a theory for anomalous high frequency
oscillations (Bose peak phenomenon)  was formulated 
within MCT \cite{mayr_bose}.  The molecular mode coupling theory (MMCT)
which is under study for a few years now, intends to extend this line
of research to experimentally relevant molecular systems. There are three
different mode coupling theories for the description of different
aspects of molecular degrees of freedom. In
\cite{single1,single2}  the motion of a single linear molecule in a
liquid of spherical atoms  is studied. In \cite{chong} a site - site
description is formulated, which is perfectly adapted to study 
neutron scattering experiments of molecular systems. In this approach
the atomic structure of the molecules is considered.
Finally, the MMCT \cite{foot2}, we are using in this work \cite{schteger}, 
is devised to 
investigate the dynamics of a liquid of linear molecules. For this
purpose a self consistent mode coupling theory for the dynamic
correlation functions of tensorial
densities $\rho_{lm}(\vec{q},t)$  was developed. These densities are the
generalized Fourier components  
of the microscopic density $\rho(\vec{x},\Omega)$ in an expansion in
spherical harmonics with respect to the orientation $\Omega = (\theta,
\varphi)$  of the
molecules  and plane waves with respect to $\vec{x}$.  An extension
to arbitrary molecules is given in \cite{wasser_test2}. A theory for
arbitrary molecules was also formulated in \cite{kawasaki}. First results for 
the tensorial nonergodicity parameters (NEP) and critical amplitudes
were obtained for dipolar
hard spheres \cite{schteger} and \cite{latzvigo}, respectively. 
A study of the phase diagram for glass transitions 
of a liquid of hard ellipsoids  was performed in
ref. \cite{ellips2}. Several aspects of
the theory for general molecules were tested against simulation for 
water in ref. \cite{wasser_test1,wasser_test3} and treating water as
a linear molecule in ref. \cite{wasser_test2}. 

As in the MCT for simple liquids the static structure factors
$S_{ll'}^m(q)$ in the q
frame, i.e. the coordinate system in which the $z$  axis is along the
wavevector,   
 completely determines the
longtime dynamics and thus the NEP and critical amplitudes. Note that
the structure factors (static and dynamic) are diagonal in $m$ in the q
-frame \cite{schteger}.  The static
structure factors have to be known to solve the equations of
MMCT. They are either obtained by analytical theories as e.g. in
ref. \cite{ram91,ram94,ellips1} for ellipsoids or they have to be taken from
simulations. 
In this work we present a detailed comparison of MMCT calculations of the NEP
and critical amplitudes with the results of simulation for a liquid of
diatomic molecules.
A description
of the system and the simulation can be found in
\cite{kammerer1,kammerer2}.  There also  tests of the
universal properties of MCT are presented. 
For our comparison between simulation and theory the static structure
factors are taken from the simulation. In a 
preliminary study only the diagonal static correlators $S_{ll}^m(q)$ 
were used as input and also the dynamical correlators $S_{ll}^m(q,t)$
and thus the NEP 
were assumed to be diagonal \cite{theis-vigo}.   
This severe approximation
has lead to unphysical results like the existence of two different
transition temperature for ODOF and TDOF.  In our study we use
diagonal {\it and} non diagonal elements of the static structure factor as
input to calculate {\it all} components of the NEP and, in addition, of
the critical amplitudes. For the calculation of the NEP,  we also
extend the necessary upper cut off $l_{co}$ for the index $l$ to
$l_{co} =4$. 

The paper is organized as follows. 
In chapter (\ref{mct_chap}) we review the main equations and concepts for
the calculation of the critical NEP (\ref{neptheory}) and the critical
amplitudes (\ref{amplitudetheory}). In chapter (\ref{results}) we
discuss the influence of different approximation schemes on the
theoretically obtained critical temperatures (\ref{tc}) and its
relation to simulation results. Then we present the comparison of
theoretical critical NEP (\ref{nepresult}) and critical amplitudes
(\ref{amplituderesult}) with simulations. Conclusions are presented in
chapter \ref{conclusions} and an appendix describes how the critical
NEP can be obtained from the liquid side close the the ideal glass
transition.

\section{Molecular Mode Coupling Theory} \label{mct_chap}
\subsection{Non ergodicity parameter}\label{neptheory}
The derivation of the equation of MMCT for the dynamics of linear molecules and
general molecules can be found in \cite{schteger} and \cite{wasser_test2}
respectively. We only repeat the basic
definitions and equations and refer the reader for details to the literature. 
For the present work, in which we want to calculate the critical 
NEP and the critical amplitudes of diatomic molecules, we only need
the equations for linear molecules in the limit of time to infinity. The
basic quantities are  the correlation functions of tensorial densities  
$\rho_{lm}(\vec{q})$ and tensorial current densities
$j^\alpha_{lm}(\vec{q},t)$   
\begin{eqnarray}
\label{I2-1}
\rho_{lm}(\vec{q},t)&=&\sqrt{4\pi} \,\, i^l \,
\sum_{n=1}^{N}e^{i\,\vec{q}\cdot\vec{x}_n(t)} 
\,Y_{lm}(\vec{\Omega}_n(t)) \quad \mbox{.}\\
\vec{j}^{\,\alpha}_{lm}(\vec{q},t) &=& 
\sqrt{4\pi}\,i^l\,\sum _{n=1}^{N} \vec{v}_n^\alpha \,
e^{i\,\vec{q}\cdot\vec{x}_n(t)} \, Y_{lm}(\vec{\Omega}_n(t)) \; .
\end{eqnarray}
The  $Y_{lm}(\vec{\Omega}_n(t))$ are the standard spherical harmonics
and we follow in our notation the text book by Gray and
Gubbins \cite{gray}. $\vec{v}_n^\alpha$ is either the center of mass
velocity $\vec{v}_n(t)$ of the $n^{th}$ molecule or its angular velocity 
$ \vec{\omega}_n(t)$ depending on the index $\alpha$: 
\begin{equation}
\vec{v}\,^\alpha _n(t) := \left\{
\begin{array}{r@{\quad,\quad}l} \vec{v}_n(t) & \alpha = T \\
\vec{\omega}_n(t) & \alpha = R \end{array} \right..
\end{equation}
where $T$ and $R$ stands for translational and rotational part,
respectively.  
For the calculation of the NEP we in principal need all spatial 
components of the currents but here we use as in \cite{schteger} only
the projection on directions
defined by the wave vector $\vec{q}$ and the angular momentum operator
$\vec{L}$. Taking into account also transversal currents will lead
only to small correction in the NEP \cite{theis00}. 
We therefore define the {\em longitudinal} currents
$j^\alpha_{lm}(\vec{q},t)$.
\begin{equation} 
j^\alpha_{lm}(\vec{q},t)=\frac{1}{q_l^\alpha}\left(\hat{q}^\alpha
\,\vec{j}^\alpha \right)_{lm}(\vec{q}) \quad , \quad \alpha \in\{T,R\}\;
\label{I2-13}
\end{equation}
with
\begin{equation}
\label{I2-14}
q^\alpha _l(q) := \left\{
\begin{array}{r@{\quad,\quad}l} q \qquad & \alpha = T \\ \sqrt{l(l+1)} &
\alpha = R \end{array} \right.
\end{equation}
and the definition 
\begin{equation}
\label{I2-15}
\hat{q}^\alpha := \left\{
\begin{array}{r@{\quad,\quad}l} \vec{q} & \alpha = T \\ \vec{L} & \alpha =
R \end{array} \right.
\end{equation}.
The quantities we are going to calculate are NEP $F^m_{ll'}(q)$ in the
q -frame i.e. in a coordinate system
in which the z-axis is given by the direction of the wave vector
$\vec{q} = (0,0,q)$. In this coordinate system all correlation functions
$\langle\rho^*_{lm}(\vec{q},t) \rho_{l'm'}(\vec{q},0)\rangle = \delta_{mm'}
S_{l,l'}^m(q,t)$  are diagonal in $m$. They are real and depend on
$|m|$ only. The same holds for all other tensorial quantities, we will
use  in the q - frame. The NEP 
are given by  
\begin{equation}
F^m_{ll'}(q) = \lim_{t\to \infty} S_{ll'}^m(q,t).
\end{equation}

As input for the mode coupling equations we need the static
structure factors $S^m_{ll'}(q)$  
They are directly taken from the simulation of K\"ammerer 
et. al. \cite{kammerer2}. The off diagonal elements of the  static
structure factors were not yet published and had to be determined from
the raw data.  

The equation for the matrix ${\bf F}(q) \equiv (F_{ll'}^m(q))$ of NEP
can then be  written \cite{schteger} as

\begin{equation} \label{frac1}
{\bf F}(q) = \left[ {\bf 1}+\mbox{\boldmath ${\cal K}$}(q) \,
{\bf S}^{-1}(q)\right]^{-1}{\bf S}(q) \quad .
\end{equation}
  
The matrix {\boldmath $\cal K $} is related to the Laplace transform of the 
dynamic current correlation function 
and can be expressed as the inverse 
of a memory matrix $\mbox{\boldmath $\cal F$}^m(q) \equiv ({\cal
  F}_{ll'}^m(q))$ at $t=\infty$,
\begin{equation}
\label{frac2}
{\cal K}^m_{ll'}(\vec{q}) = \sum_{\alpha
\alpha '} q_l^\alpha (q)
\left(\mbox{\boldmath ${\cal F}^m$}(q)^{-1} \right)
_{ll'}^{\alpha \alpha '}q_{l'}^{\alpha '}(q).
\end{equation}
 The mode coupling approximations yield:  
\begin{eqnarray}
\label{mct}
& & \!\!\!\!\!\!\!\!\!\!\!\! {\cal F}_{ll '}^{m\;\alpha \alpha '}(q)\approx
\frac{1}{2N} \left ( \frac{\rho_0}{4 \pi}\right ) ^2
{\sum _{\vec{q}_1 \vec{q}_2}}' \sum_{m_1m_2}\sum_{l_1l_2}\;\sum_{l_1'l_2'}
\times \nonumber \\
& & \times \, V^{\alpha \alpha '}_{ll'l_1l_1'l_2 l_2'}(q,q_1,q_2 ; m, m_1, m_2)
\, F^{m_1}_{l_1 l_1'}(q_1)\, F^{m_2}_{l_2l_2'}(q_2) \; ,
\end{eqnarray}

with 

\begin{eqnarray}
\label{I2.3-q5}
& & \!\!\!\!\!\!\!\! V^{\alpha \alpha '}_{ll'l_1l_1' l_2 l_2 '}
(q , q_1,q_2 ; m, m_1, m_2) :=  \nonumber \\
& & v^{\alpha}_{ll_1 l_2} (q, q_1,q_2;m, m_1, m_2)\cdot
v^{\alpha '}_{l' l_1  'l_2 '} (q, q_1  , q_2; m, m_1, m_2 )^*\quad ,
\end{eqnarray}
\begin{eqnarray}
\label{vertex}
& &  \!\!\!\!\!\! v^{\alpha}_{l l_1 l_2} (q,q_1,q_2; m, m_1, m_2) :=
 \nonumber \\
& & \sum _{l_3} u^{\alpha}_{l l_3l_2}(q , q_1  , q_2; m, m_1, m_2 )\,
c_{l_3l_1}^{m_1}(q_1)+ (-1)^m(1\longleftrightarrow 2)
\end{eqnarray}
and 
\begin{displaymath} 
u^{\alpha}_{l l_1
l_2}(q , q_1  , q_2; m, m_1, m_2 ) :=
i^{l_1+l_2-l}\left[\frac{(2l_1+1)(2l_2+1)}{(2l+1)} \right]^{\frac{1}{2}}
\frac{1}{2}\left[1+(-1)^{l_1+l_2+l}\right] \times
\end{displaymath}
\begin{eqnarray}
& \times & \sum_{m_1'm_2'}(-1)^{m_2'}d_{m_1'm_1}^{l_1}(\Theta_{q_1})
\,d_{m_2'm_2}^{l_2}(\Theta_{q_2})\,C(l_1l_2l;m_1'm_2'm) \nonumber \\
\label{mctend}
& \times & \left\{\begin{array}{l@{\quad;\quad}l}
q_1\cos\Theta _{q_1}\,C(l_1l_2l;000) & \alpha = T \\
\sqrt{l_1(l_1+1)}\,C(l_1l_2l;101) & \alpha = R \end{array} \right. \quad
\mbox{.}
\end{eqnarray}

Here the functions $C(l_1l_2l;m_1,m_2,m)$  are the Clebsch Gordan
coefficients and $d_{m'm}^{l}(\Theta)$ are 
related to
Wigner's rotation matrices (we follow the notation of Gray and
Gubbins). For given Euler angles $\Omega=(\Phi,\Theta,\chi)$ 
they are defined  as
\cite{gray})  
\begin{equation}
\label{I2.3-2}
D_{mm'}^l(\Omega)=e^{-im\Phi}\,d^l_{mm'}(\Theta)\,e^{-im'\chi}.
\end{equation}

$\Theta_{q_i}$ is the angle between $\vec{q}$ and $\vec{q}_i$. 
The prime at the first summation in Eq.(\ref{mct}) restricts
$\vec{q}_1,\vec{q}_2$ such, that $ \vec{q}_1+\vec{q}_2 =\vec{q}$. 
Eqs. (\ref{frac1} - \ref{frac2}) and (\ref{mct}) form a closed set of
infinitely many coupled nonlinear equations for the NEP. To obtain a
solvable theory we 
have to restrict $l$ to be smaller than an upper cut off, $l \le
l_{co}$. The resulting equations can in principle 
be solved by a fixed point iteration algorithm. The
physical control parameter like the temperature and the
density only enter via the static structure factor. At  a critical
temperature or density the solution of this equations bifurcates from
all functions $F_{ll'}^m(q)$ being zero to nonzero. In the simulations
of K\"ammerer et al. \cite{kammerer1,kammerer2} the temperature was
used as control parameter. 
Close to the
transition temperature $T_c$ the stability matrix of the iteration
(see below) will have 
a largest eigenvalue $E_0$ approaching $E_0=1$ from below.  Consequently the
convergence of the iteration is getting very  slow close to $T_c$. The 
time for one iteration depends very sensitive on the upper cut off
$l_{co}$. For $ l_{co}=2$ one iteration took $10$ minutes on a MIPS
R10000, for $
l_{co}=4$ this time increased to 6 hours. We therefore
concentrated on $l_{co} =2$, to determine the transition point with a 
high accuracy and used the calculation for $l_{co} =4$ mostly as check 
for the sensitivity of our results against changing the cut off. 

To overcome some of the restrictions connected to the critical slowing
down of the convergence close to $T_c$ we determined the critical NEP,
i.e. the NEP at $T_c$, with two
alternative methods. For the standard fixed point iteration we started  at
a temperature low enough to be in the glass state. Then the
temperature is increased very slowly. At every temperature the
equations for the NEP are solved by  the iteration 

\begin{equation} \label{iterate}
{\bf F}^{(n+1)}(q) = \mbox{\boldmath $\cal G$}({\bf F}^n,\epsilon) 
\end{equation}
where $\mbox{\boldmath $\cal G$}({\bf F}^{(n)})$ is the right hand side of
Eq. (\ref{frac1}) and $\epsilon = (T_c -T)/T_c$ . This iteration
converges exponentially fast towards 
its solution as long as the temperature is not the critical
temperature. The convergence rate is determined by the largest
eigenvalue $E_0$ of the stability tensor ${\cal  C}_{\lambda
  \lambda'} = \partial {\cal G}_\lambda/\partial {
  F}_{\lambda'}$ \cite{foot3}. The index $\lambda$ is an abbreviation for
wavevectors $q$ and rotational indices $l,l',m$. The exponential
convergence rate is then $\ln E_0$. Close to and below $T_c$ the
eigenvalue $E_0$ can
be written as $E_0= 1 - A \sqrt{\epsilon} $ with A being a positive
constant.  
Therefore the convergence rate  is $A  \sqrt{\epsilon}$ and the number of
iterations to obtain convergence diverges inversely proportional to
$\sqrt{\epsilon}$  close to $T_c$. With this number of iterations the
deviations of our NEP from the true critical NEP are proportional
to $\sqrt{\epsilon}$, since the NEP exhibit the well known square
root singularity (cf. \cite{leshouches}). 

If the temperature is increased above $T_c$, i.e. $\epsilon <0$,
there is no  non zero solution  for the iteration
Eq. (\ref{iterate}). 
Nevertheless for $0 < -\epsilon \ll 1$, the iteration is
nearly stationary for  a large number of iteration of the order
$|\epsilon|^{-1/2}$ (see appendix). The approximate critical NEP
is determined as the stationary point ${\bf \hat{F}}(\epsilon)$
whose change along the 
eigenvector with eigenvalue 1 of the {\em critical} stability
matrix is minimal during iteration. 
But contrary to the NEP determined from the fixed point iteration
for $T <T_c$, the stationary NEP differ only in order $|\epsilon|$ instead
in order $\sqrt{\epsilon}$ from the true critical NEP. Consequently
this property allows us 
in the following to crosscheck the very accurate results for $l_{co} =2$
obtained from the fixed point iteration and also to obtain the critical NEP for
$l_{co}=4$.

\subsection{Critical amplitudes}\label{amplitudetheory}
A central prediction of the mode coupling theory of the glass
transition in simple liquids is the existence of the $\beta$ -
relaxation regime \cite{go84,go85,leshouches}.  For the problem of a
single dumbbell in an isotropic hard sphere system \cite{single1} 
it was demonstrated, that the $\beta$ relaxation law can be detected in every
quantity, which couples  to the density. For a liquid of anisotropic
molecules it is not very well defined which degree of freedom is driving the
glass transition. The equations of MMCT couple all degrees of freedom
and there are situations where the transition is not caused by the
TDOF, but the ODOF \cite{ellips2}. But even in these systems the
factorization theorem is generically valid for all correlators. This 
can be proven using the 
standard techniques \cite{leshouches}. Therefore every dynamic
structure factor $S_{ll'}^m(q,t;T)$
can for $ -1 \ll \epsilon \ll 1$ in the
$\beta$ relaxation regime  be written as 
\begin{equation} \label{beta}
S_{ll'}^m(q,t;T) = F_{ll'}^m(q) + H_{ll'}^m(q) G(t/t_0; \sigma)
\end{equation}

The function $G(t/t_0)$ is  the same for all $l,l',m,q$. $t_0$ is an
overall microscopic scale. 
$H_{ll'}^m(q)$ are the critical amplitudes determining the intensity
of the asymptotic $\beta$ - relaxation for a certain combination of
$l,l',m$ and $q$.  
Also the correction to the asymptotics, which determine, besides
the temperature, the range of validity of the law Eq. (\ref{beta}) 
depend on these amplitudes (cf. ref \cite{corrections} for simple liquids).
Differences in the observability of the critical correlators between
depolarized light scattering experiments and dielectric loss
measurements, 
can be explained by differences in the amplitudes $H_{ll'}^m(q)$
involving $l=2$ and $l=1$ respectively (see ref. \cite{single2} for a
single molecule). To determine the amplitudes
numerically to a high precision it is necessary to be very close to the
transition point, to make sure that all correction terms of order
$\epsilon$ are small compared to the leading term of order
$\sqrt{\epsilon}$. Due to the difficulties described above we could
only
determine the critical amplitudes for upper cut off $l_{co}=2$.

\section{Results}\label{results}
\subsection{The critical temperatures \label{tc}} 
For $l_{co}=2$ and in the full diagonalization approximation
\cite{theis-vigo}, in which the 
static structure  factors $\Sq$, the glass form factors $\Fq$ and the
memory matrix $\Mq$ are assumed to be diagonal with respect to $l$, the glass
transition temperature for the TDOF predicted by MMCT is below the transition
temperature of the MD simulations $T_c^{MD} = 0.477$. Note, that this
temperatures are given in Lennard Jones units (cf. \cite{kammerer1}). 
In all other known examples the
MCT overestimates the tendency for vitrification. 
As an additional artefact of the full diagonalization the ODOF vitrify
at a lower temperature than the TDOF. 
Since the top down symmetry of the dumbbells is broken, the full
equations of MMCT (\ref{frac1}) - (\ref{mct}) do generically
not allow for such a scenario.  
As soon as we take $\Sq$ non diagonal all degrees of freedom
undergo a glass transition at the same temperature above the MD
result. To study the influence of different diagonalization
approximations a bit
more in detail, we investigated several cases, with the main condition
of $\Sq$ being non diagonal:

\begin{enumerate}
\item $\Fq$ and $\Mq$ diagonal \label{dd} \qquad \mbox{(dd)} 
\item $\Fq$ diagonal and $\Mq$ non diagonal \label{dnd} \qquad
  \mbox{(dnd)} 
\item $\Fq$ and $\Mq$ non diagonal \label{ndnd} \qquad \mbox{(ndnd)} 
\end{enumerate}

\begin{figure}[tbp] 
\centerline{\epsfxsize=8cm \epsfysize=7cm
\epsffile{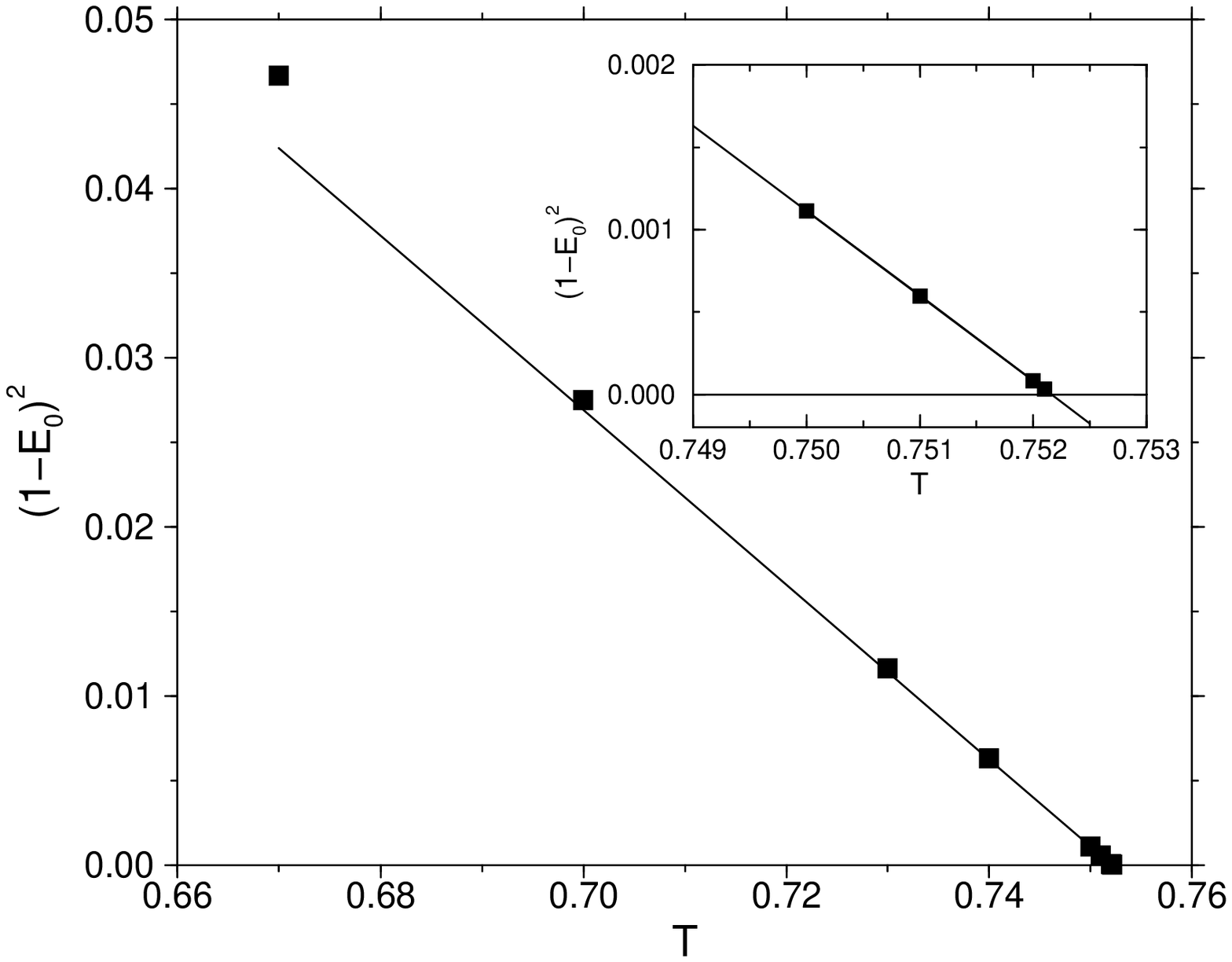}}
\caption[E0]{
The square deviation of the largest Eigenvalue of the stability matrix (see
text) from its critical value
$E_0(T_c)=1$  as function of temperature. The full line is a linear
fit $(1- E_0)^2 \propto (T_c - T)^2 $ to the
data with $T_c=0.75215$. The inset shows a magnification for the same
quantity very close to $T_c$. } 
\label{E0} 
\end{figure}

\begin{figure}[tbp] 
\centerline{\epsfxsize=8cm \epsfysize=7cm
\epsffile{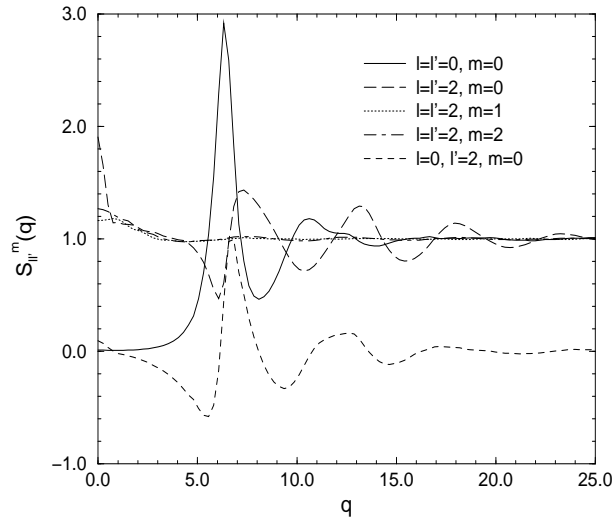}}
\caption[fig1]{
A selection of structure factors $S_{ll'}^m(q)$ of a liquid 
of diatomic molecules
in the q -frame at temperature $T=0.7521$ and pressure $p=1$ in
Lennard Jones units. 
The full curve is the structure factor $S^0_{00}(q)$, the long dashed curve
is $S^0_{22}(q)$,the dotted
curve is $S^1_{22}(q)$,  the dashed dotted curve is $S^2_{22}(q)$ 
and the short dashed  curve is  $S^0_{02}(q)$}. 
\label{fig_sqlm} 
\end{figure}

Let us discuss $l_{co}=2$ first. For this case $T_c$ has been determined
 very accurately from the asymptotic behavior $(1- E_0(T))^2 \propto
 \epsilon$ (see chapter \ref{amplitudetheory}) for the largest
 eigenvalue $E_0(T)$. Fig. \ref{E0}. demonstrates this law for case
 (\ref{ndnd}).  
The highest transition temperature is obtained for case
(\ref{dd}). Here the transition temperature is roughly three times as
large as the MD result, $T_c^{dd} = 1.4$. In case (\ref{dnd}) it only slightly
decreases to $T_c^{dnd} = 1.38$. If everything is taken non diagonal
(case \ref{ndnd}) the transition temperature is $T_c = 0.7521$.
Although still twice as large as the MD result, the discrepancy of this
result from $T_c^{MD}$ is comparable to other known cases  and
consistent with the
usual $20\%$ accuracy of the critical {\em density}
\cite{kob_nauroth}. The equations are too complex to get a deeper
theoretical understanding of this seemingly erratic jumping of the
transition temperature, dependent on the approximation we are
using. Particularly the fact, that the vertices (Eq. \ref{vertex}) are not
positive anymore, makes an analytical prediction impossible. 
But it is at least possible to rationalize the behavior using a
combination of physical and mathematical arguments. First of all it is
quite clear that the full diagonalization where all matrices are
assumed to be diagonal is too crude to describe the coupling of TDOF
and ODOF for the system of diatomic molecules.  The TDOF and ODOF are
only coupled via the diagonal memory function ${\cal F}_{ll
'}^{m\;\alpha \alpha'}(q)$. 
The coupling of the equations
for different $l$ is considerably reduced compared to the case where
$\Sq$ is taken to be non diagonal. 
E.g. in the 
${\cal F}_{00}^{m\;T T}(q)$ component of the memory
matrix only terms of the 
form (symbolically) 
$\sum_{l',m'}V^{m'}_{l'}  (F^{m'}_{l' l'})^2 $  appear in the full
diagonalization approximation, since the Clebsch Gordan coefficients
$C(0l'l'',0,m',m'')$, which enter into the vertices (cf. Eqs. (\ref{mct})
- (\ref{mctend}))  are nonzero for
$l'=l''$ only. Similarly, the memory functional
${\cal F}_{11}^{m\;\alpha \alpha'}(q)$  only contains couplings of the
form $ F^0_{00} 
F^m_{11}$ and $F^m_{22} F^m_{11}$. For ${\cal F}_{22}^{m\;\alpha
\alpha'}(q)$  
the Clebsch Gordan coefficients in the vertex allow ``self'' 
couplings $(F^m_{11})^2$, $(F^m_{22})^2$ and couplings to $l=0$ in the
form $F^0_{00} F^m_{22}$, but no ``self'' couplings
$(F^0_{00})^2$. Due to the absence of $(F_{00}^0)^2$ in 
${\cal F}_{22}^{m\;\alpha \alpha'}(q)$  a freezing of the center of
mass motion, i.e. $l=l'=0$ does 
not imply a freezing for quadrupolar dynamics $l=l'=2$. 
If the vertex for the coupling of the NEP with $l=2$ and 
$l=0$ in $F^m_{11}$ and
$F^m_{22}$ are not large
enough, exactly this structure of the memory matrix allows generically a
separate transition  of the $l=0$ and the $l \ne 0$ components of the
diagonalized dynamic structure factor as observed in \cite{theis-vigo}. But
we have to stress that the approximation is not per se inadequate. In
the case of water \cite{wasser_test1} the full diagonalization approximation
leads to a rather satisfactory agreement with simulations, without the
artefact of separate transitions and too low transition
temperatures. The pronounced angular dependence of interaction between
water molecules,
which is reflected in the fact that the static structure
factors for $l=2,m=0,1,2$ and $l=0$  are of the same order, yield large enough
vertices to produce a single transition temperature of the TDOF and
ODOF. In the present case, the structure factor $S^0_{00}$ is clearly more
dominant than $S^m_{22}$ (see Fig.  \ref{fig_sqlm}). 
This is different from water where all $m$ for $l=l'=2$ are
important.
This leads to the a posteriori conclusion that in general the full
diagonalization can 
only be used (if at all) for systems with ``very strong'' static translation
rotation coupling. This statement, unfortunately, cannot be further quantified.

If we now take the non diagonality of $\Sq$ serious 
but leave all other matrices diagonal (case \ref{dd}),  {\em additional}
coupling between TDOF and ODOF appear, which may lead to an
effectively stronger coupling. Although the equations for the
different $l$ components of the NEP still couple only via the diagonal
memory functions ${\cal F}_{ll}^{m\;\alpha \alpha'}(q)$, the 
diagonalized memory matrix contains now {\em static} couplings between
{\em all}
NEP. E.g., ${\cal F}_{11}^{m\;\alpha \alpha'}(q)$  contains  
additional couplings   between
the TDOF - correlator $F^0_{00}$ and the correlators involving $l=2$
and, even more important, "self coupling" terms $(F_{00}^0)^2$ due to the
non vanishing structure factors $S^0_{10}$ and $S^m_{12}(q)$,
respectively. This of course does
not explain, but at least makes plausible the dramatic increase of the
transition temperature. Any slowing down of let's say the TDOF is
immediately transferred to all other degrees of freedom and causes a
further slowing down of the TDOF due to the feed back via the memory
function. This enhances the tendency towards vitrification and also is
responsible for the existence of a single transition temperature.  

The reason for the decrease of the transition temperature, when  we give
up the diagonalization approximations for $\Mq$ and $\Fq$ (case
\ref{ndnd}), is not obvious. 
We only note, that the transition temperature is decreasing from case
\ref{dd} ($T_c = 1.4$)  over \ref{dnd} ($T_c=1.38$) to \ref{ndnd}
($T_c = 0.752$) i.e. the more off diagonal
elements of the matrix \mbox{\boldmath $\cal F$} are taken into account. 
\mbox{\boldmath $\cal F$} is the $t \to \infty$ limit of the
memory matrix i.e. the random force correlation function.  Therfore it
seems, that the more components of the random forces are coupled, the
lower is the transition temperature. This implies, that the
more the random forces can mutually influence each
other, the more difficult it
is to form a glass.  Although we cannot proof this statement on mathematical
grounds, it describes an feasible physical phenomenon.

To test the sensitivity of our results to changes in the cut off, we
also solved the MMCT for upper cut off $l_{co}=4$. The larger cut off
value for $l$ reduces the transition temperature further towards the
simulation result. For $l_{co}=4$, an upper and a lower bound for
$T_c$ has been  determined. The lower bound is the highest temperature
for which the NEP are still nonzero after 88 iterations.  The upper
bound is the temperature for which the NEP are converging to zero
after about 24 iterations. 
Since the time per iteration increases dramatically
upon increasing the upper cut off the transition temperature could
only be determined within $5\%$, $T_c = 0.61$. It is encouraging that
the real transition is approached upon increasing the cut
off, but our arguments presented above, show, that this is  not
necessarily the case. Which of the competing mechanisms influencing
the transition temperature is dominant, cannot be predicted on general
grounds.    
 
\subsection{The non ergodicity parameters \label{nepresult}}
\begin{figure}[tbp] 
\centerline{
\epsfxsize=8cm \epsfysize=7cm \epsffile{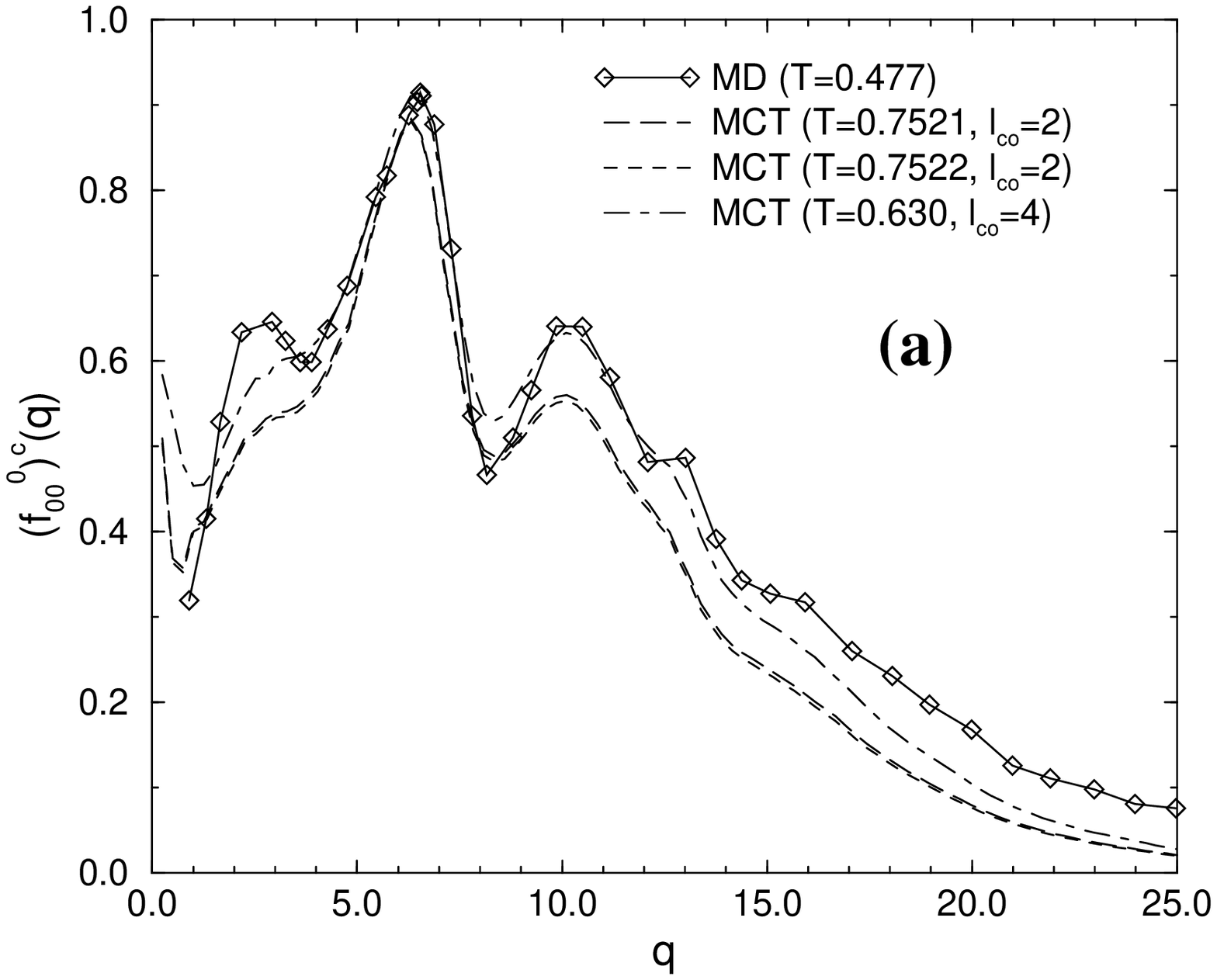} 
\epsfxsize=8cm \epsfysize=7cm \epsffile{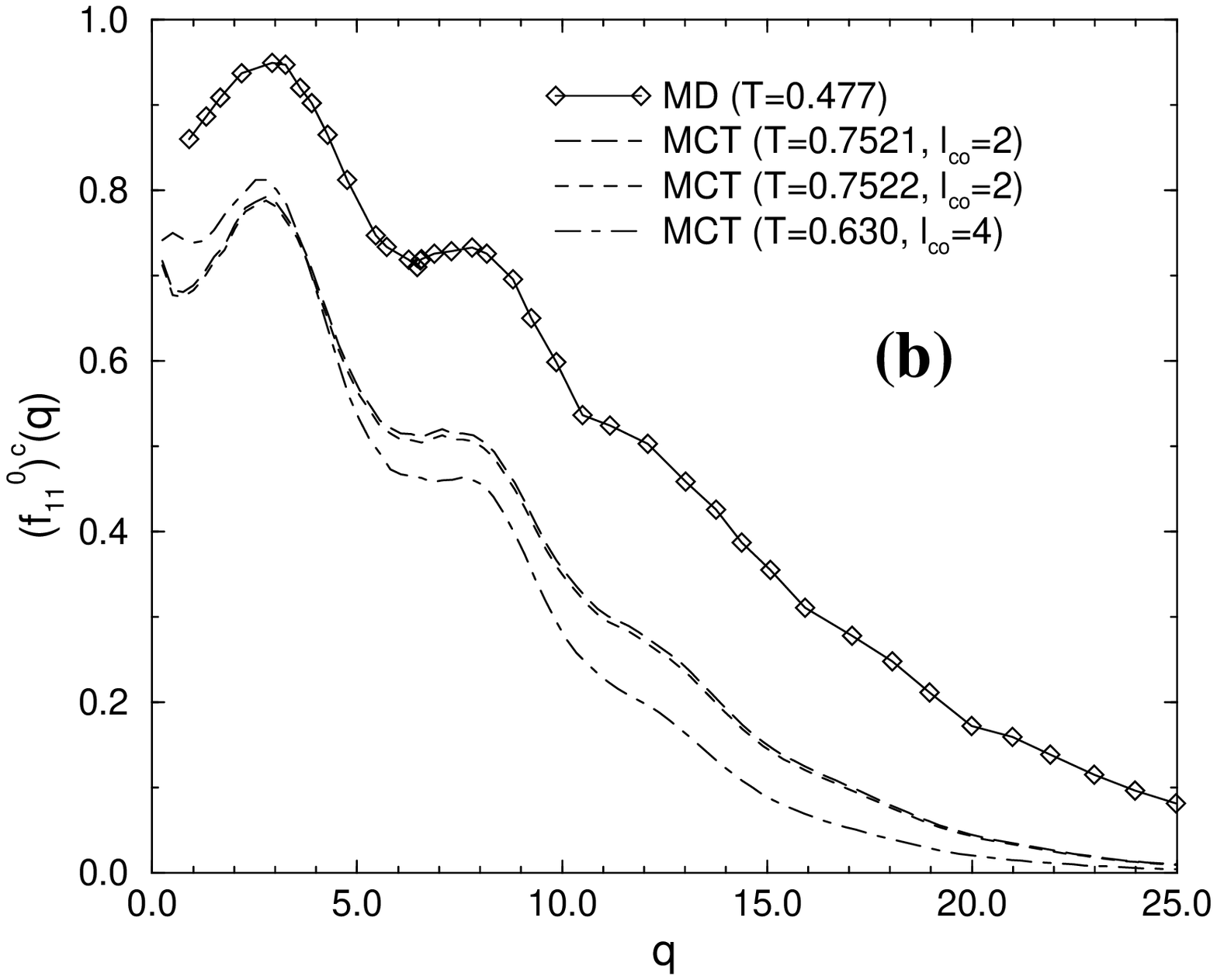}}
\centerline{
\epsfxsize=8cm \epsfysize=7cm \epsffile{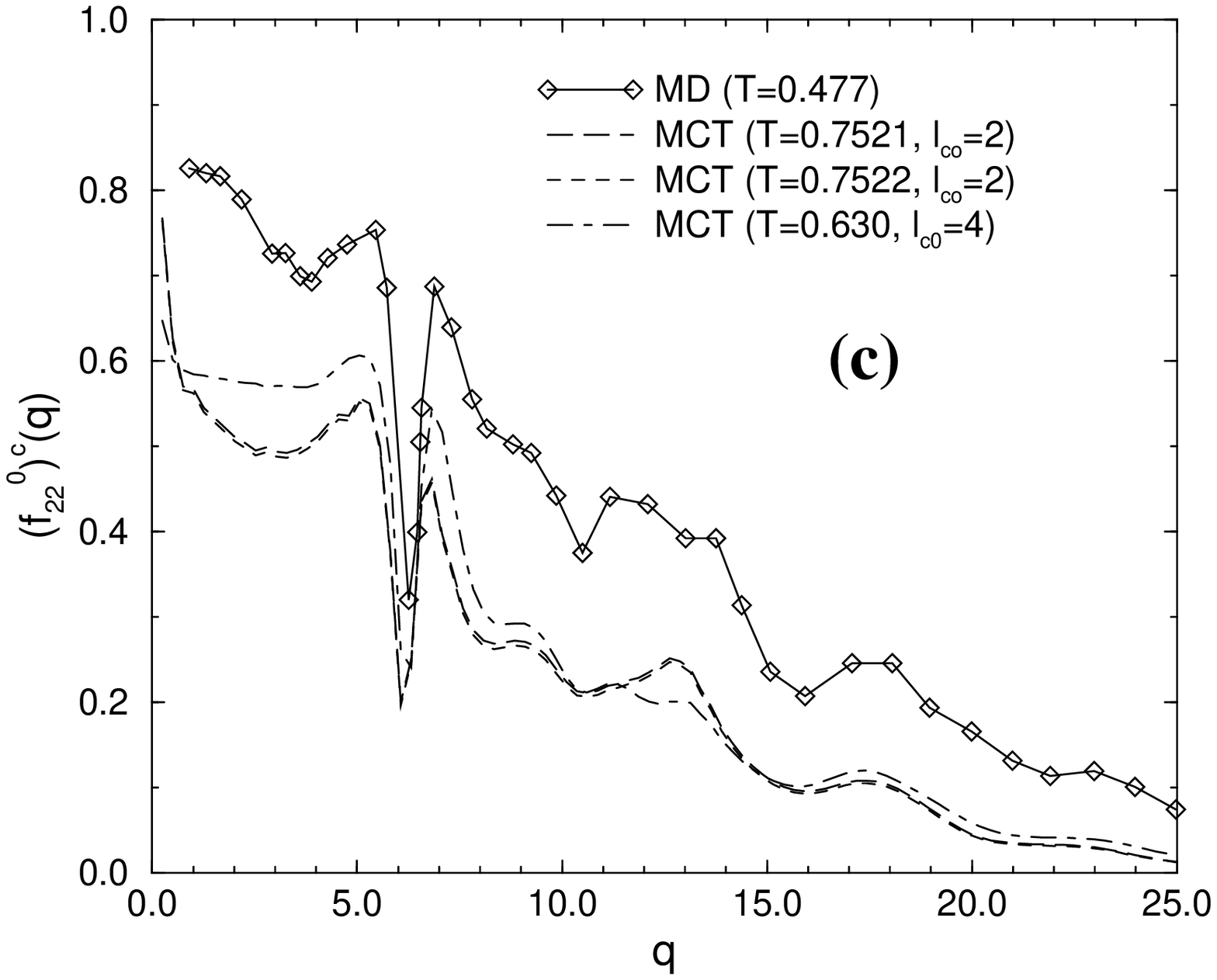} 
\epsfxsize=8cm \epsfysize=7cm \epsffile{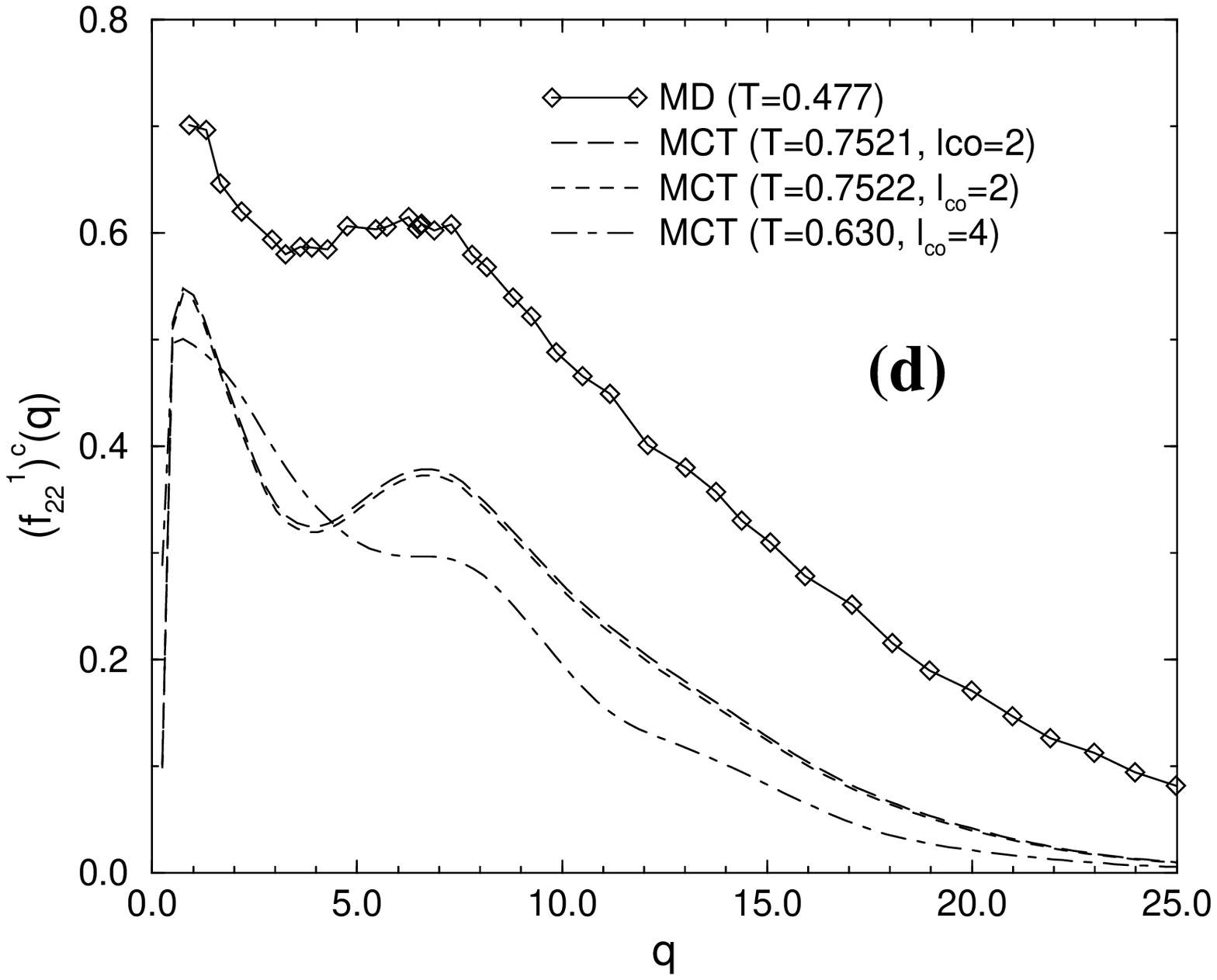}}
\caption[fig2]{Normalized diagonal elements $f_{ll}^m(q)$ 
 of the non ergodicity parameter matrix. As normalization the
 corresponding diagonal elements of the static structure factor are
 used. Shown are the elements for $(l,m) = (0,0)$ (a) ,$(l,m) =(1,0)$
 (b), $(l,m) =(2,0)$ (c)  and $(l,m) =(2,1)$ 
(d). The full line is the result of the simulation
at $T=0.477$ obtained by fitting a von Schweidler law plus corrections
of order $(t/\tau_{\alpha})^{2b}$ to the simulation data. The
estimated critical temperature is $T_c=0.475$.
 The long dashed line is the result of the fixed point
iteration for upper cut off $l_{co}=2$ at $T = 0.7521$ slightly below
the theoretical $T_c=0.75215$. The short dashed curve is the result of the
quasi stationary 
criterion for upper cut off $l_{co}=2$ at $T= 0.7522$ slightly above
the theoretical $T_c$. The dashed dotted curve is the result of the
quasi stationary
criterion for upper cut off $l_{co}=4$ at $T= 0.630$ (above the
theoretical $T_c$) for $l_{co}=4$.}  
\label{fig_nepdiag} 
\end{figure}

In the following we concentrate on the results for the normalized NEP 
$f_{ll'}^m(q) = F_{ll'}^m(q)/\sqrt{S_{ll}^m(q) S_{l'l'}^m(q)}$ without any
diagonalization approximation. Fig. \ref{fig_nepdiag}  shows the
normalized diagonal
terms of the matrix of NEP $f_{ll}^m(q)$ for 
$(l,m) = (0,0),(1,0),(2,0),(2,1)$. 
Not shown are the results for $(l,m) = (1,1), (2,2)$, since they do not exhibit
very much structure.  The corresponding simulation result is taken from
\cite{kammerer2}. It was obtained by fitting a von Schweidler law plus
corrections 
$S_{ll'}^m(q,t) = F_{ll'}^m(q) - H_{ll'}^m(q) (t/\tau_\alpha)^b +
(H^{(2)})_{ll'}^m(q) (t/\tau_\alpha)^{2b}$ to the simulation results
for the time dependent density correlation function $S_{ll'}^m(q,t)$,
where $\tau_\alpha$ is the $\alpha$ - relaxation scale.   
There are three different  
theoretical curves.  The two curves at temperatures $T_c=0.7522, 0.7521$ are
obtained with the fixed point method (on the glass side of the
transition) and the quasi stability criterion (on the liquid side), 
respectively, as described
above for upper cut off $l_{co}=2$.  Their good agreement demonstrates
the high accuracy  of the
solution. The third theoretical curve shows the result for upper cut
off $l_{co}=4$ using the more accurate quasi stability criterion. 
Compared to the results in \cite{theis-vigo} a clear improvement
of the agreement with simulations can be observed. Especially the q -
dependence 
of the functions are very well reproduced. Even a feature like the
prepeak in $f^0_{00}(q)$ at $q \sim 2.5$ is reproduced as a shoulder in
the corresponding 
theoretical result. 
This peak is not present in the static structure factors. Since
$S_{11}^0(q)$ has a peak at about $q \sim 2.5$ it could appear due to
a dynamic coupling of the ODOF, especially the one
involving $l=1$ and the TDOF. Note also, that the mentioned peak exactly
corresponds to the first peak in $f_{11}^0(q)$.    
There is a tendency, that  
the agreement is best around the wave vector, 
where the structure factor $S_{00}(q)$ has its first peak 
 and is getting
worse for large wave vectors. This might be
interpreted as an indication for the glass transition being  driven also for
the investigated system of diatomic molecules by the TDOF. From
investigations of other systems \cite{schteger,latzvigo,ellips2}, we know that
different scenarios are possible.  

\begin{figure}[tbp] 
\centerline{
\epsfxsize=8cm \epsfysize=7cm \epsffile{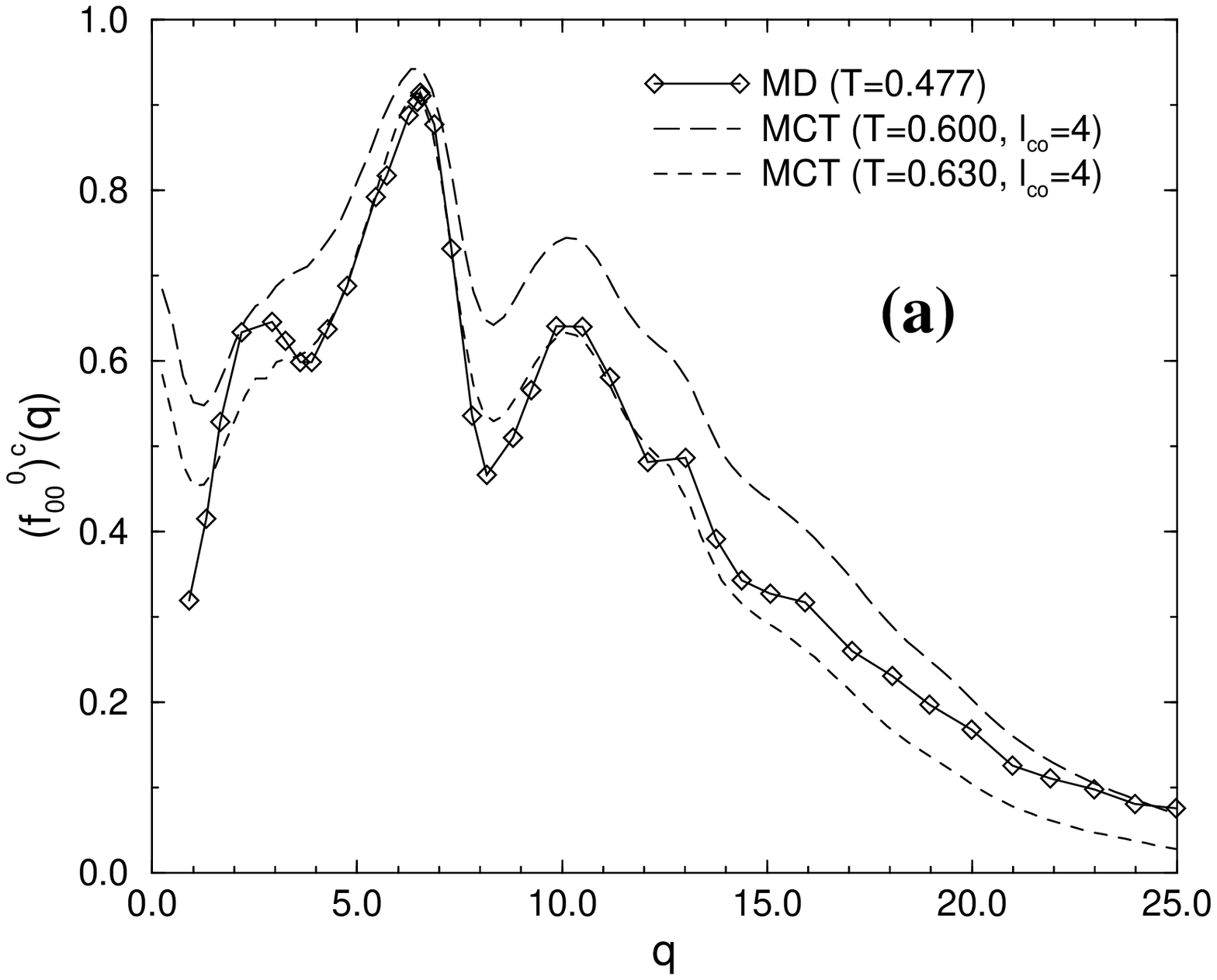} 
\epsfxsize=8cm \epsfysize=7cm \epsffile{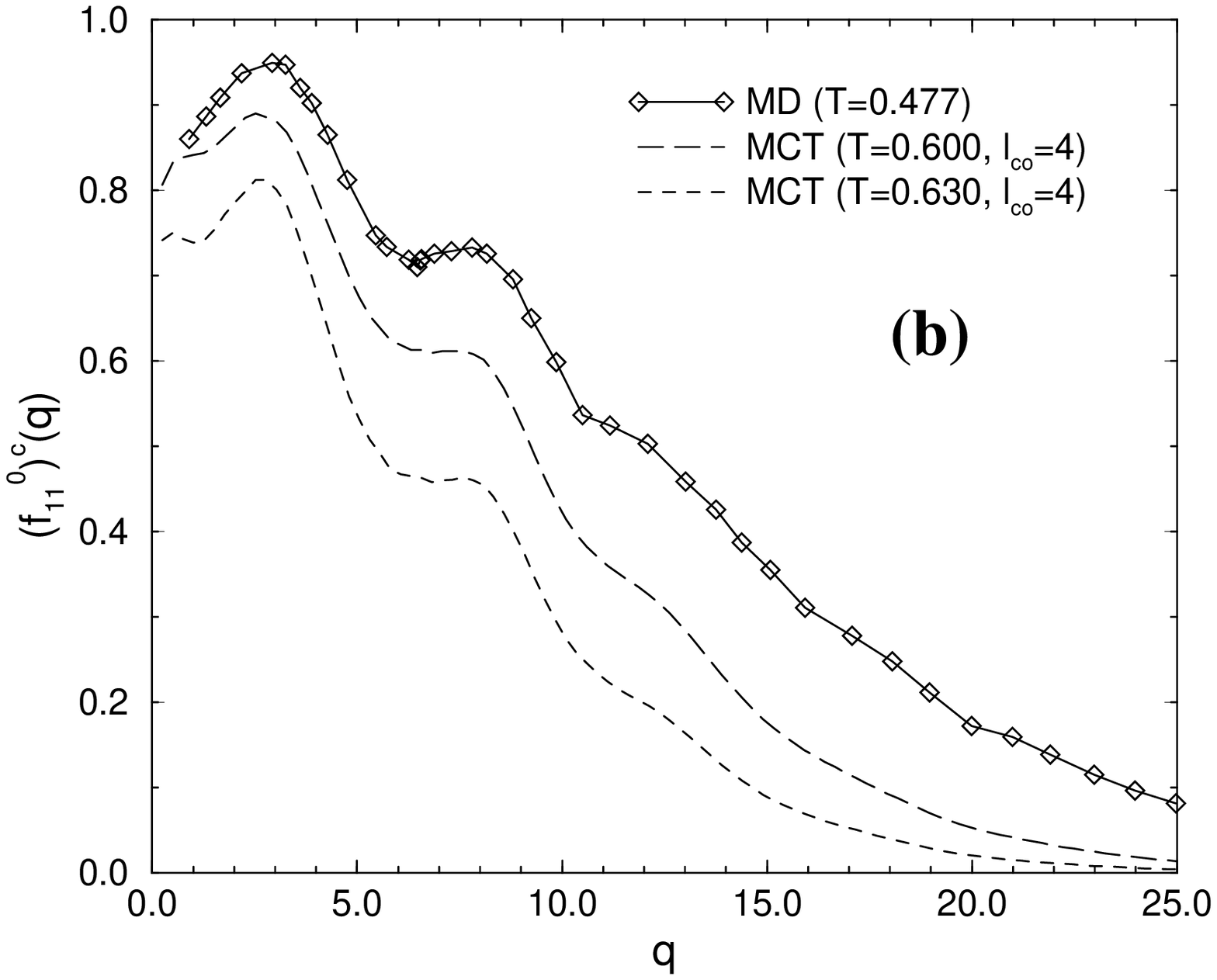}}
\centerline{
\epsfxsize=8cm \epsfysize=7cm \epsffile{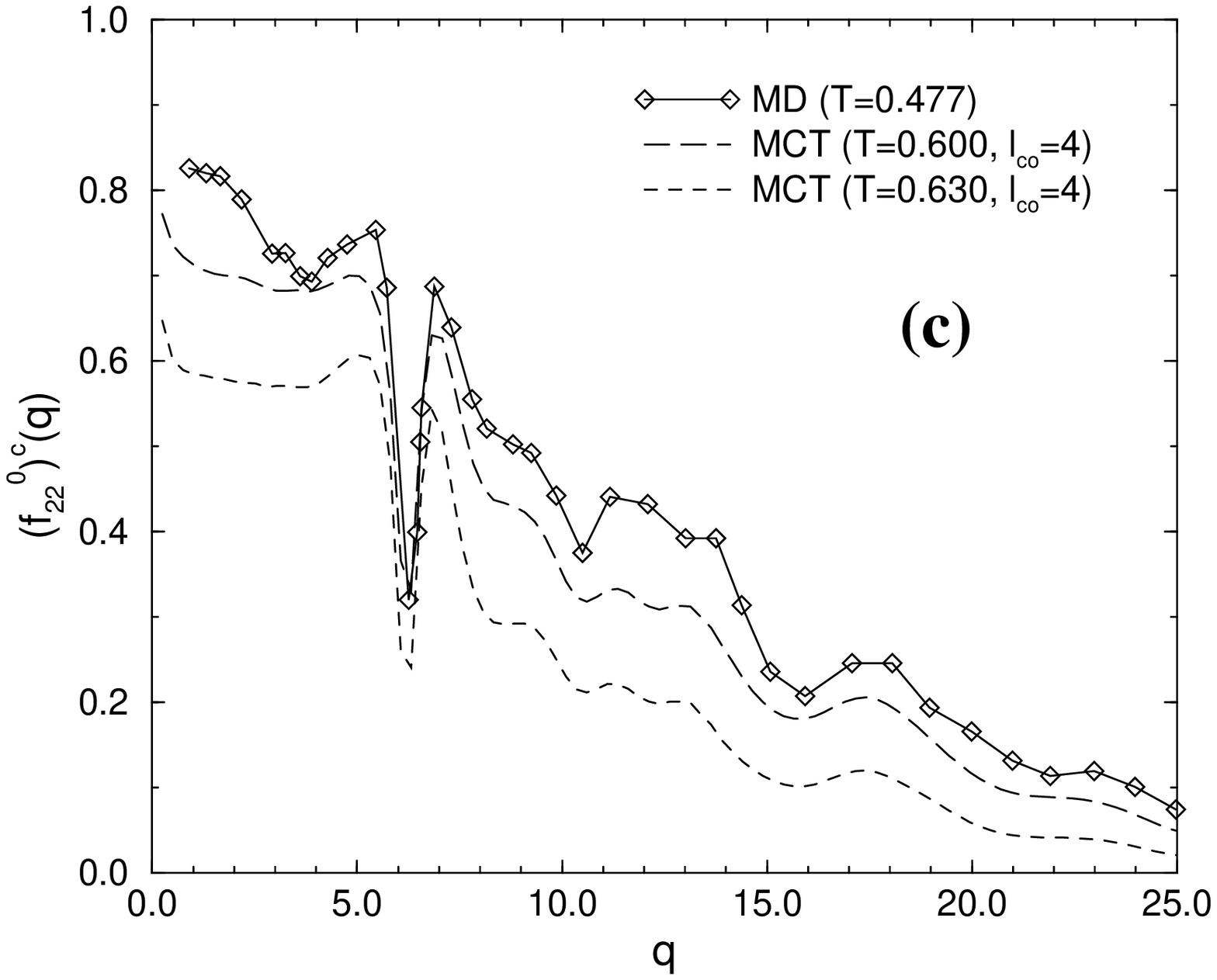} 
\epsfxsize=8cm \epsfysize=7cm \epsffile{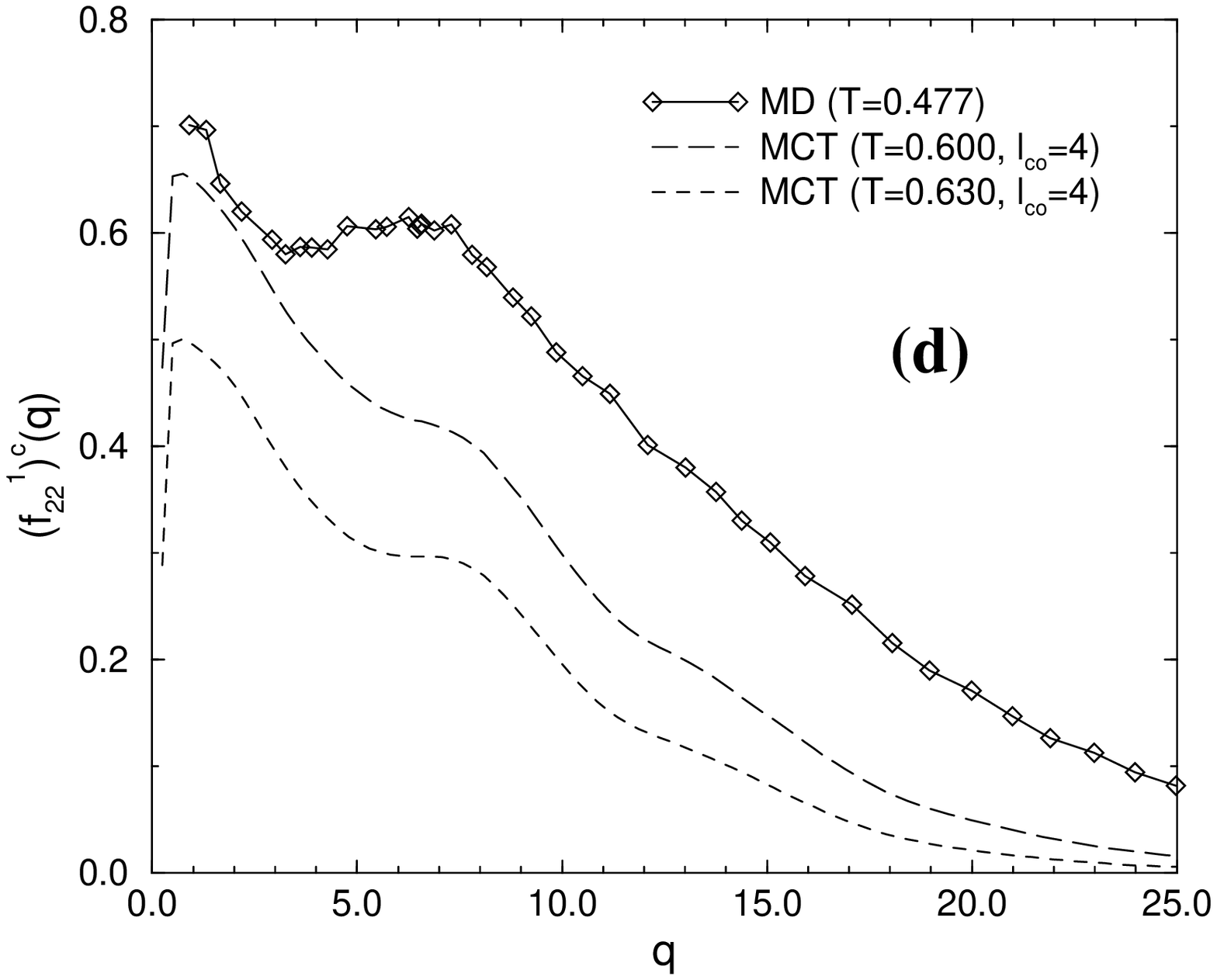}}
\caption[fig3]{
Normalized diagonal elements $f_{ll}^m(q)$ 
 of the non ergodicity parameter matrix for $l_{co}=4$ obtained with
 fixed point method and stationary criterion respectively in comparison
 with simulation.
 As normalization the
 corresponding diagonal elements of the static structure factor are
 used. Shown are the elements for $(l,m) = (0,0)$ (a) ,$(l,m) =(1,0)$
 (b), $(l,m) =(2,0)$ (c)  and $(l,m) =(2,1)$ 
(d). The full line is the result of the simulation
at $T=0.477$.
 The long dashed line is the result of the fixed point
iteration for upper cut off $l_{co}=4$ at $T = 0.600$. 
The short dashed curve is the result of the quasi stationary 
criterion for upper cut off $l_{co}=4$ at $T= 0.630$ }
\label{fig_lco4} 
\end{figure}

Similar to increasing $q$, the agreement between simulation and theory gets
worse with increasing $l$. This is expected due to two different
reasons. First higher $l$ correspond to a higher angular resolution and
are therefore probably much more affected by the mode coupling
approximation.  Second, higher $l$ are of course much more sensitive
to the cut off $l_{co}$ than lower $l$. The curves for larger cut off
increase the quality of the comparison with the MD results. But it is
important to note that in our case no general rule can be given of
how much the quality of the results for lower values of $l$ can be
improved by increasing the upper cut off, as this was done in
\cite{single1}. In the case of a single dumbbell in a liquid of
hard spheres the glass transition temperature is completely determined
by the hard sphere liquid and does not change by increasing
the cut off $l_{co}$. As explained above, in our case, $T_c$ can depend
very sensitively on $l_{co}$. But this influences directly the
amplitude of the NEP via the trivial effect of the temperature on the
static structure factors. We already compensate as much as possible for
this mechanism by presenting only the {\it normalized} NEP. But as in
the case of hard spheres, there is still the effect, that also the
normalized NEP are proportional to the static structure factor. This
is a very nontrivial phenomenon, since the existence of negative
vertices in the mode coupling functional $\mbox{\boldmath ${\cal F}$}(q)$
could in principle lead to a violation of this correlation. 
But as can be
inferred from Fig. \ref{fig_nepdiag} the NEP $f_{00}^0(q),
f_{22}^0(q)$ for $l_{co}=4$ are
systematically larger  than the one for $l_{co}=2$ in a large region
around the first peak of the 
structure factor $S_{00}^0(q)$ , without big
differences in the functional form. This effect, especially for
$F_{00}^0(q)$ where the mentioned trend is valid for all wavevectors, 
 can be mainly
understood  as a consequence of the transition temperature being
smaller for $l_{co}=4$ than for $l_{co}=2$, which causes the 
the first peak of $S_{00}^0(q)$ to increase.
Additional evidence for this reasoning is presented in
Fig. \ref{fig_lco4}. In this figure we show the
results for $l_{co} =4$ at temperature $T=0.60$, obtained with fixed point
method, and $T= 0.63$, obtained with the quasi stability criterion. The
lower temperature is still in the glass (The required accuracy i.e. the
maximum difference between consecutive  iterates is smaller than
$10^{-6}$, 
is reached after 88 iterations). At the higher temperature the
accuracy is first increasing as expected (see appendix). But after 24
iterations it begins slowly to decrease and after 42 iterations the
iterates start to converge quickly towards the solution $\Fq=0$. 
The results for the lower
temperature agree, except for $l=0$,  much better with the simulation, 
than the one for
the higher temperature. But since the deviations to the true critical
NEP at roughly the same number of iterations are of order $|T-T_c|$
for the higher temperature compared to
order $|\sqrt{|T-T_c|}$ for the lower one, we have to
conclude, that the results for the higher temperature are closer to the
critical NEP of the theory. The better agreement with simulations 
of the NEP at $T= 0.600$ is a trivial consequence of the fact, that
positive NEP increase with decreasing temperature. 
Due to this influence of the value of $T_c$ on the amplitude of the
normalized critical NEP, we cannot in general conclude that increasing
the cut off $l_{co}$ leads to a better agreement with the simulation.
It might even happen, that  the agreement with simulations gets worse
instead of better, if increasing the cut off would lead to a larger
transition temperature. This is possible due to the existence of
negative vertices in the mode coupling functional {\boldmath{$\cal F$}}. 

The observed trends allow the reasonable hypothesis, that the
temperature effect could
be the main source for the deviations
between simulation and theory. In general the main structural features
in the normalized non ergodicity parameter are very well
represented, but they are systematically to small for nearly all
wave vectors, exactly as expected, if the theoretical transition
temperature is too large.

\begin{figure}[tbp] 
\centerline{\epsfxsize=8cm \epsfysize=7cm \epsffile{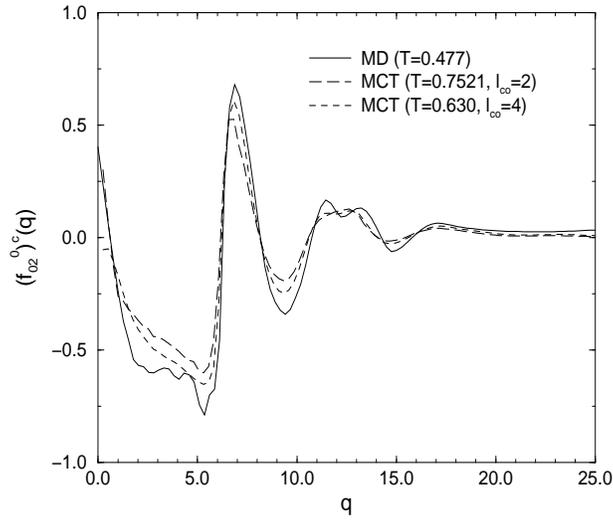}}
\caption[fig4]{
Normalized off diagonal elements $f_{02}^0(q)$ 
 of the non ergodicity parameter matrix.  
 As normalization $\sqrt{S^0_{00} S_{22}^0}$ was chosen. 
The full line is the result of the simulation
at $T=0.477$.
 The long dashed line is the result of the fixed point
iteration for upper cut off $l_{co}=2$ at $T = 0.7521$. 
The short dashed curve is the result of the quasi stationary 
criterion for upper cut off $l_{co}=4$ at $T= 0.630$.} 
\label{fig_nondiag} 
\end{figure}

Fig. \ref{fig_nondiag} demonstrates, that the theory also gives good results
for the off diagonal NEP. We found that $S_{02}^0$ and $F_{02}^0$ are the only
important off diagonal components of the static structure factor matrix
$\Sq$ and NEP matrix ${\bf F}(q)$, respectively.  
In Fig. \ref{fig_nondiag} we therefore show the normalized NEP $f_{02}^0$. 
The quality of the result is even better, than for the
diagonal components of the NEP.   

\subsection{Critical amplitudes}\label{amplituderesult}
\begin{figure}[tbp] 
\centerline{
\epsfxsize=8cm \epsfysize=7cm \epsffile{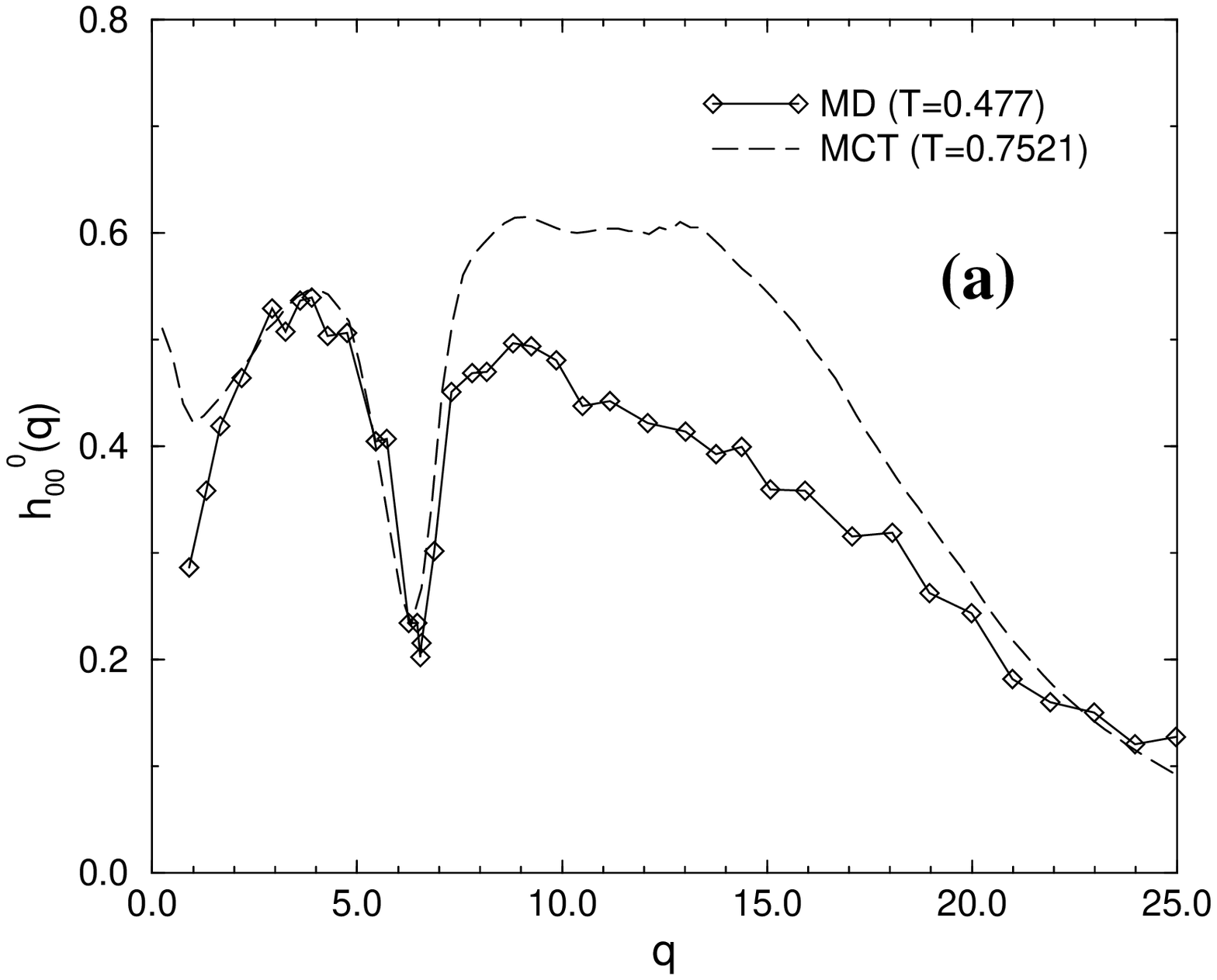} 
\epsfxsize=8cm \epsfysize=7cm \epsffile{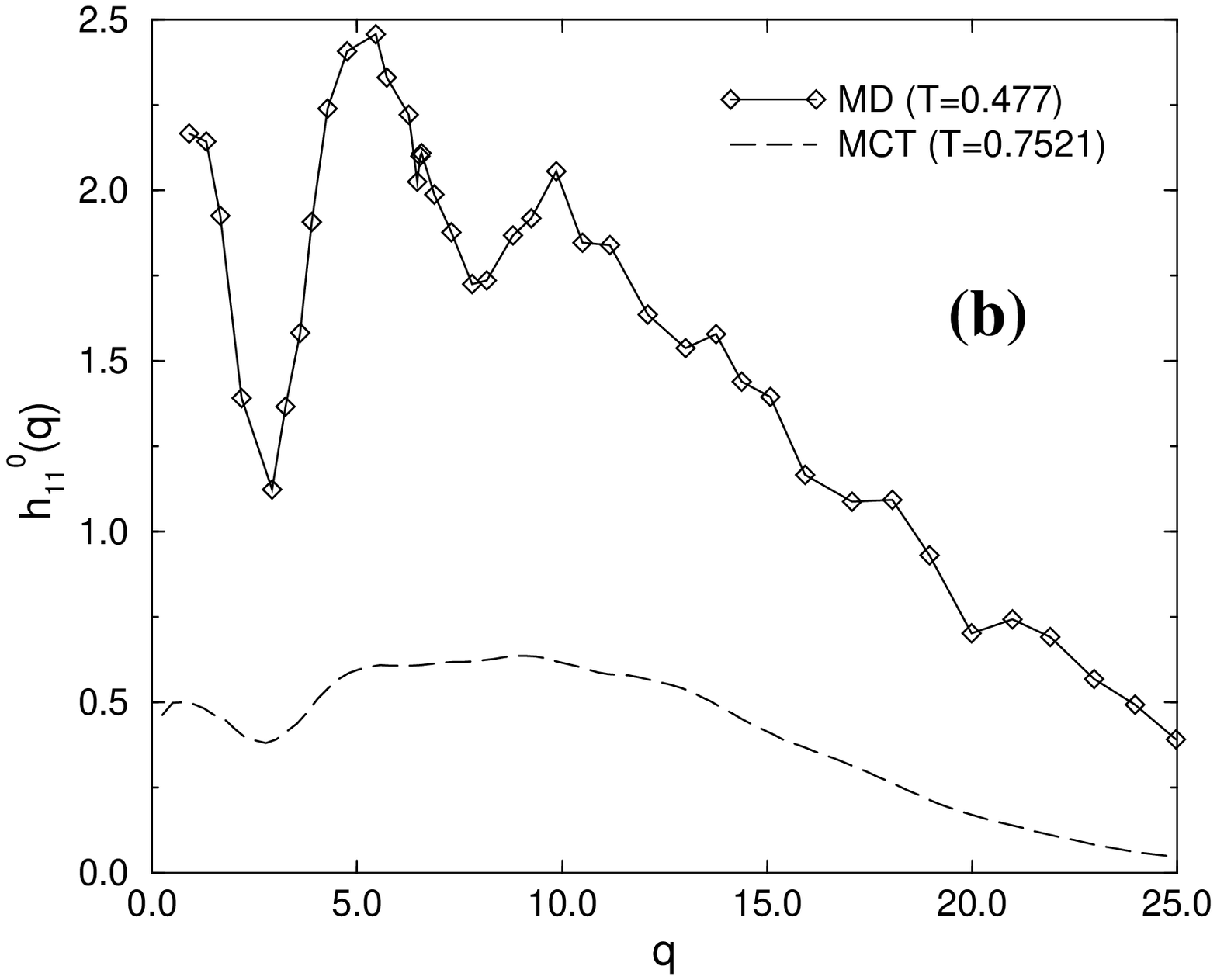}}
\centerline{
\epsfxsize=8cm \epsfysize=7cm \epsffile{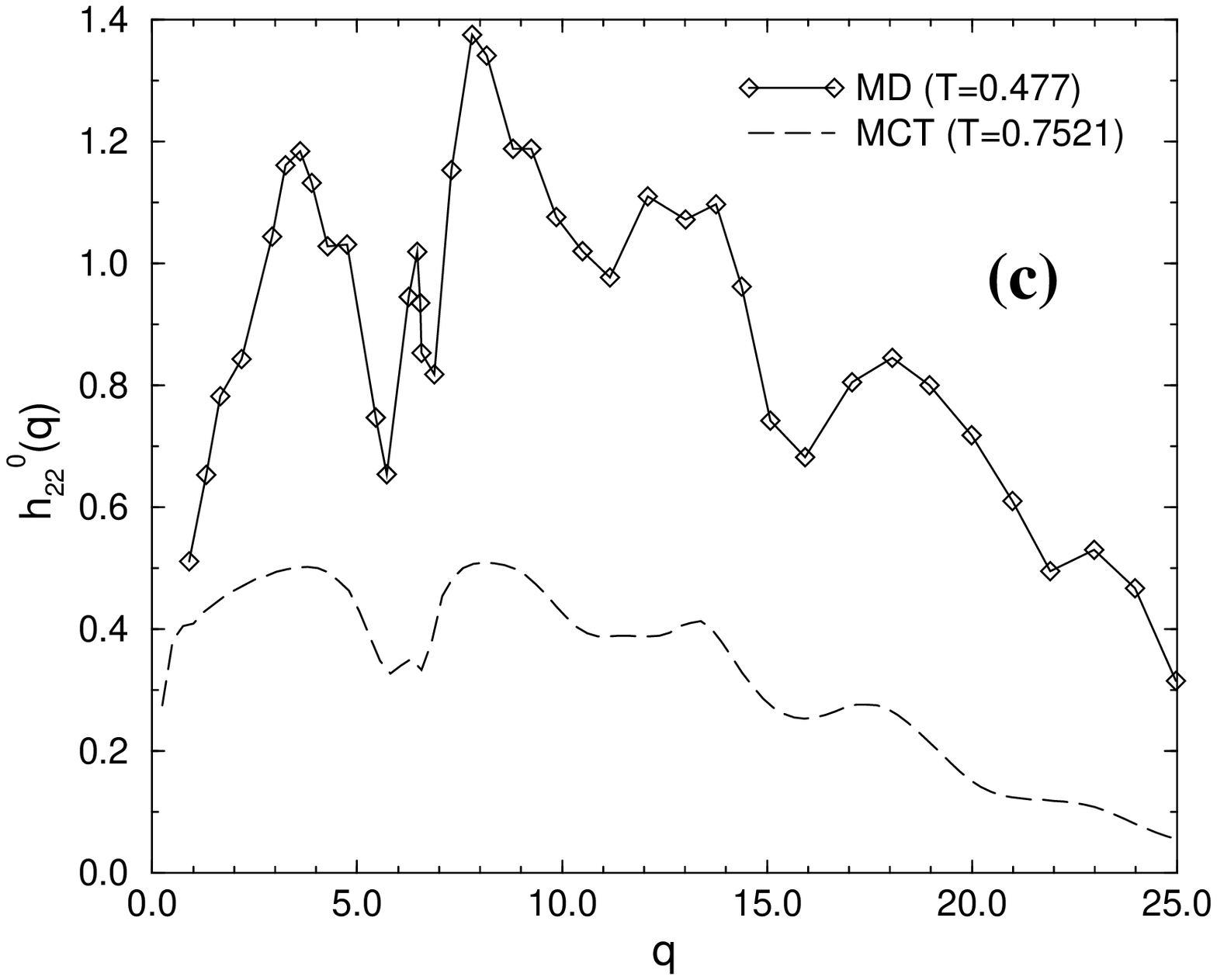} 
\epsfxsize=8cm \epsfysize=7cm \epsffile{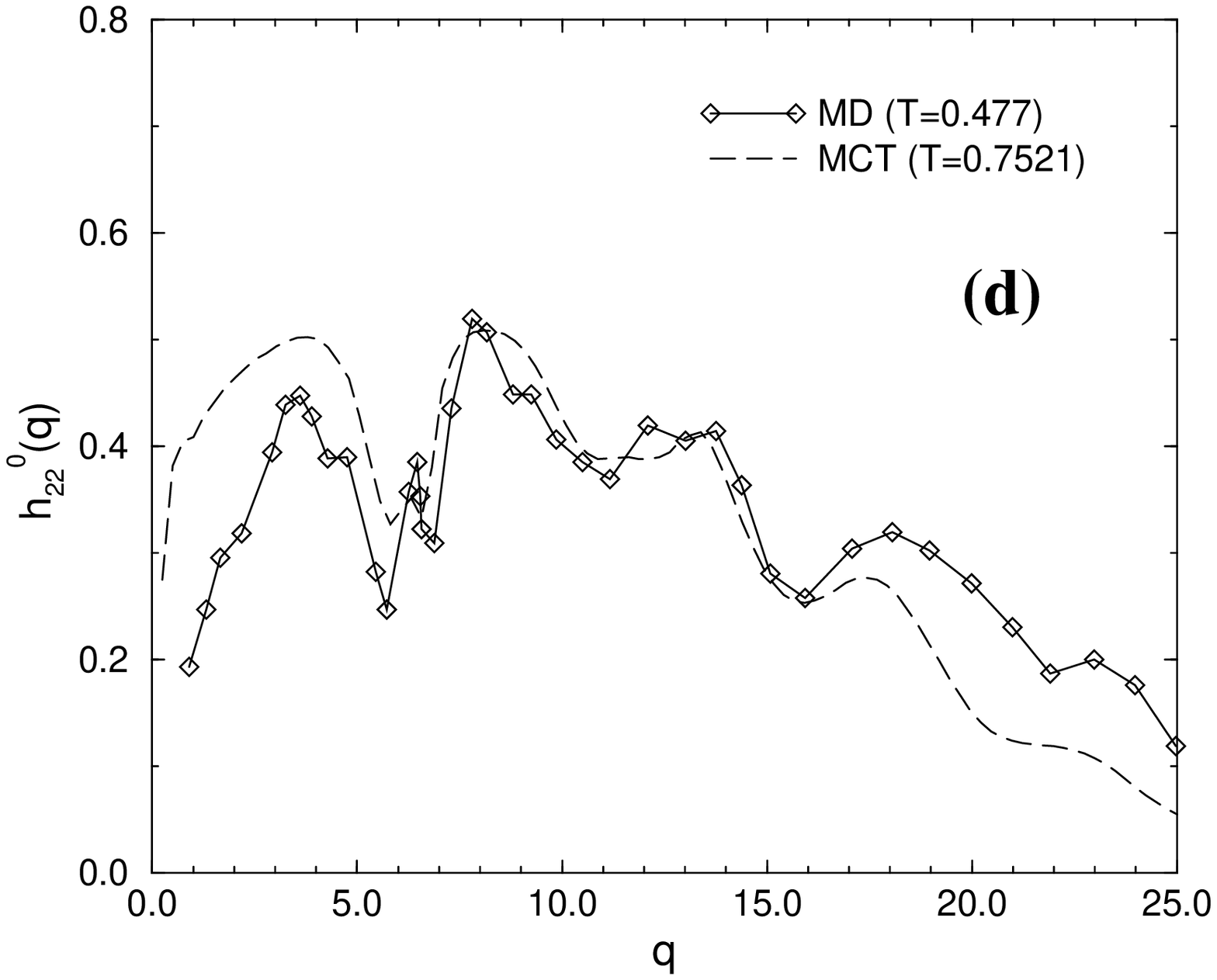}}
\caption[fig5]{Normalized diagonal elements $h_{ll}^m(q)$ 
of the matrix of critical amplitudes.  As normalization the
 corresponding diagonal elements of the static structure factor are
 used. The result of the simulation
at $T=0.477$ is obtained by fitting a von Schweidler law plus corrections
of order $(t/\tau_{\alpha})^{2b}$ to the simulation data. 
The fits to the simulations (full lines) are multiplied with a
factor 200 (see text for details) to obtain best agreement with
the theoretical result for  $h_{00}^0(q)$.  
Shown are the theoretical results for $T=0.7521$, upper cut off
$l_{co}=2$. \ref{fig_hqlm}(a)-(c) show the results for $(l,m)
 = (0,0),(1,0),(2,0)$ respectively. The dashed  curve is always
the result of the theory. \ref{fig_hqlm}(d) shows again the same
theoretical result as in Fig \ref{fig_hqlm}(c), but  the free 
scale factor the critical
amplitude of the simulation is chosen to optimize the agreement with
theory (see text for details).} 
\label{fig_hqlm} 
\end{figure}

The critical amplitudes are determined only up to an overall scale
factor. I.e. our theoretical results cannot be directly compared to
the simulation results. But once we have chosen a scale factor for
e.g. the amplitude $h^0_{00}$, all other amplitudes should be
multiplied with the same scale factor to compare with the
simulations. In fig \ref{fig_hqlm}(a) we have chosen a scale factor of 200 to
obtain best agreement with the normalized critical amplitude
$h_{00}^0(q)= H_{00}^0(q)/S_{00}^0(q)$. The features of this component
are the same as in
simple glass forming systems
\cite{mebagoe,colloid2,meba,kob_andersen}. 
There is a
minimum at the position of the first peak of
$S^0_{00}(q)$. Simulations and theory 
compare quite well for $2\le q \le 7$ and show deviations at other
wave vectors. With the chosen scale factors the other diagonal elements
of the critical amplitude matrix  show strong deviations from the
simulation results. Especially $h^0_{11}(q)$ (see Fig.\ref{fig_hqlm}(b)) 
does not even disagree in
amplitude but also in the form, except for the minimum at $ q \sim
3$. This is not unexpected, since the  simulations 
have shown strong differences between the form of the
dynamic correlators involving odd and even $l$ \cite{kammerer2}. 
 Due to the weak top down anisotropy of our diatomic molecules, the dynamic of
the correlators involving odd $l$ is only weakly coupled to the dynamics
of the even components \cite{single2}. $180^0$ jumps are still possible
on a much faster time scale than the translational motion
\cite{kammerer2}. 
Consequently there are strong corrections to the asymptotic results
for the correlator $S^0_{11}(q,t)$ and the amplitude $h^0_{11}(q)$ is
not very well defined. 

The deviations between simulation and theory in $h^0_{22}$ (see
Fig. \ref{fig_hqlm}(c)) are more
serious, since $S_{22}^0(q,t)$ exhibits a well defined $\beta$ - 
relaxation regime. We can improve the agreement between simulation and
theory by choosing a free scale factor for the simulation curves. 
The result is shown in
Fig. \ref{fig_hqlm}(d) 
to demonstrate, in contrast to $h_{11}^0$, that essential structural
features in
$h_{22}^0$ are indeed reproduced by the theory. As argued above the dynamic
correlators and therefore also the critical
amplitudes  involving $l=2$
are  much more affected by the cut off $l_{co} =2$ as the one with
lower $l$. This might be the reason for the rather large discrepancy
found for $h_{22}^0$.
To determine the critical
amplitudes, it is necessary to be very close to the critical
point. Restriction in computer time didn't allow us to determine $T_c$
for $l_{co}=4$ with high enough accuracy to get reliable results for
the critical amplitudes.

\begin{figure}[tbp] 
\bigskip 
\centerline{\epsfxsize=8cm \epsfysize=7cm \epsffile{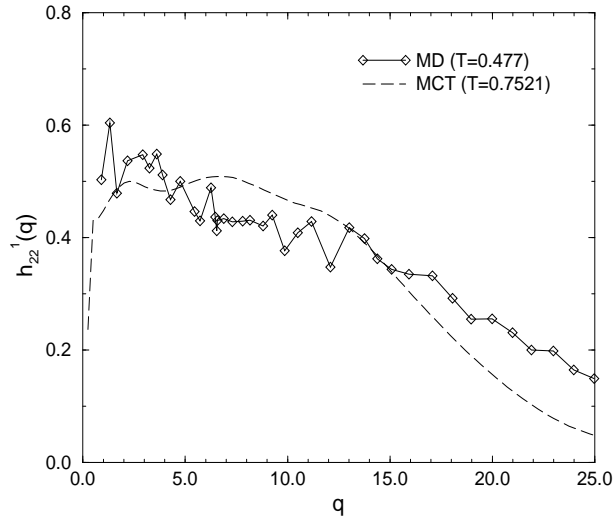}}
\caption[fig6]{Normalized critical amplitude $h_{22}^1(q)$. 
The critical amplitude of the simulation (full line) is multiplied with a
free scale  factor to optimize agreement with theory (see text for
details). Shown is the theoretical result for 
 $T=0.7521$ (dashed line).}
\label{fig_hq221} 
\end{figure}

\begin{figure}[tbp] 
\bigskip 
\centerline{\epsfxsize=8cm \epsfysize=7cm \epsffile{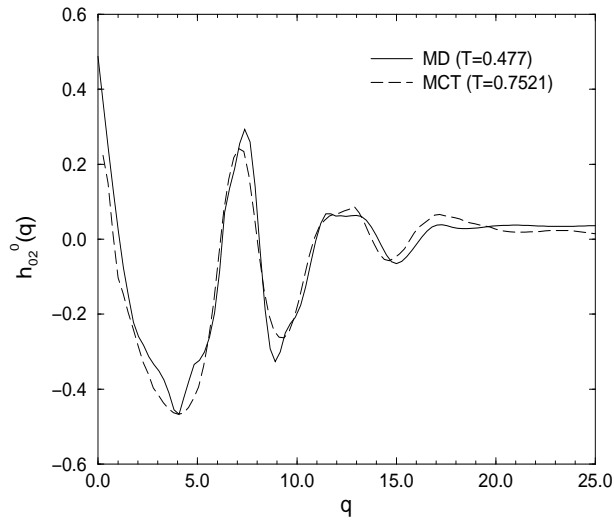}}
\caption[fig7]{Normalized critical amplitude $h_{02}^0(q)$. 
The critical amplitude of the simulation at $T=0.477$ (full line) is 
multiplied with a free scale  factor to optimize agreement with theory
(see text for details). Shown is the theoretical result for
 $T=0.7521$ (dashed line). Since the raw data of the simulations had to
be smoothed, the data points are omitted. }
\label{fig_hq020} 
\end{figure}

The critical amplitudes with $m>0$ do not exhibit very much
structure. In Fig. \ref{fig_hq221} 
we show for completeness the result for 
$h_{22}^1(q)$, which could in principle be measured in light scattering
experiments. Again we choose an overall amplitude
prefactor as a free fit parameter, but the agreement is still not
very good. Much better is the agreement (similar  to the NEP) for $h_{02}^0$
(see Fig. \ref{fig_hq020}),
although we still had to choose the amplitude scale of the simulation
as a free fit parameter.

\section{Conclusions} \label{conclusions}
We have performed a quantitative test of MMCT for a liquid of diatomic 
molecules.  The static structure factors from simulations were used as 
input for the MMCT to calculate the critical temperature $T_c$, the
matrix of critical NEP $F_{ll'}^m(q)$ and the matrix of critical
$\beta$ - 
relaxation amplitudes $H_{ll'}^m(q)$. Several approximation schemes were
used to test the sensitivity of the results against changing the
degree of diagonalization of the $F_{ll'}^m(q)$  in $l$ and $l'$ and the
dependence on the upper cut off $l_{co}$. As maximum cut off, $l_{co}=4$ was
used. Since the computational effort increases strongly with $l_{co}$, we
used a new, more accurate, method to determine the critical NEP 
from the liquid side of the transition. 

As expected, by giving up any diagonalization approximation for
$S_{ll'}^m(q)$, $F_{ll'}^m(q)$  and the memory functional ${\cal F}_{ll
'}^{m\;\alpha \alpha '}(q)$  all
degrees of freedom vitrify at a single temperature. In fact, to obtain 
a unique transition temperature, it is sufficient to keep only the static
structure factor $S_{ll'}^m(q)$ nondiagonal in $l$ and $l'$. The strongest
effect of successively applying diagonalization approximations for the
NEP $F_{ll'}^m(q)$  and the memory functional 
${\cal F}_{ll '}^{m\;\alpha \alpha'}(q)$  is the change
of the transition temperatures $T_c$. Also using different cut offs $l_{co}$
changes the transition temperature. 

Contrary to $T_c$, the overall form of NEP is much less sensitive  to the 
different approximation schemes.  The agreement with simulation is
qualitatively quite good  for cut off $l_{co}$=2 as well as for
$l_{co}$=4. In some cases especially for the $l=0$, $l'=0$ component
and the off  
diagonal $l=0$, $l'= 2$ component of the NEP the agreement is even
quantitative for wavevectors around the first peak of the structure
factor $S_{00}^0(q)$. The comparison of the NEP with simulations for the 
correlator with  ($l=0$, $l'=0$) is clearly improved using the cut off
$l_{co}=4$ instead of $l_{co}=2$. Other correlators are better represented for
$l_{co}=2$. We therefore cannot conclude that in general a further increase
of the cut off will lead necessarily to still
better agreement with simulations.   

Also the wave vector dependence of the normalized critical $\beta$-relaxation 
amplitudes agrees quite well with the 
one obtained from simulations.  Contrary to the prediction of MMCT 
no common
amplitude scale for the critical amplitudes with different $l$ and $l'$
could be found. At present it is not clear whether this failure is
real or due
to the fitting procedure used in \cite{kammerer2} to obtain the beta
relaxation amplitudes. Also the restriction to cut off $l_{co}=2$ for the
determination of the critical amplitudes could cause the found
discrepancies, since also correlators with $l>2$ will contribute to the
critical amplitudes at $l=2$. By neglecting them an error in the
amplitude scale is possible.     

In summarizing our study, we may say, that the MMCT offers an overall
consistent description of the glass transition in  molecular liquid of diatomic
molecules, at least concerning the critical NEP and the critical
amplitudes.  

\begin{appendix}
\section{Determination of the NEP using quasi  stability}
If the temperature is chosen above the critical
temperature, the only stable fixed point of the iteration (\ref{iterate})
is ${\bf F}=0$. But there is still the possibility of determining the
critical NEP with even higher accuracy than with the converging
iteration for $T < T_c$. 
To implement the method, we have to initialize the iteration with a
$F_\mu^{(0)}$, which is close 
to $F_\mu^c$, where  $\mu$ is a superindex for $q,l,l'$ and $m$.
This can be achieved  by first using the fixed point iteration
\begin{equation} \label{a1}
F^{(n+1)}_\mu = {\cal G}_\mu(\{F^{(n)}_{\mu'}\},\epsilon).
\end{equation}
below $T_c$ 
and increasing the temperatures in small
steps until the liquid regime is reached.  
The critical temperature will eventually be missed by a small
value $0< -\epsilon \ll 1$ and the iteration converges
after a {\em long} transient quickly to zero. 
 Since the iteration at a given
temperature is always initialized with the final result of the previous
temperature, the above assumption is fulfilled. The long transient is
caused by the marginal
stability of the critical point.  The iteration will be dominated for a
long time by the critical direction of the iteration, which is given
by the eigenvector $e^c$, which belongs to the eigenvalue $E_0=1$ of the 
{\em critical}
stability matrix $(C^c_{\mu \mu'}) = (\partial {\cal G}_\mu/\partial
F_{\mu'})^c$ \cite{foot3}. The NEP $\hat{F}_\mu(\epsilon)$ at which
the iteration is
almost stationary is given by minimizing the distance of the projection of 
$\Delta C_\mu = {\cal G}_\mu(\{F_{\mu'},\epsilon\}) -
F_\mu $ 
on the critical direction $\hat{e}^c$, where for technical reasons the 
{\em left} critical eigenvector $\hat{e}^c$ was chosen. 
This condition can be written as 
\begin{equation} \label{stationary}
\frac{\partial  \sum_{\mu} \hat{e}^c_{\mu} \Delta C_{\mu} }{\partial
F_{\mu'}} =  \sum_{\mu} \hat{e}^c_{\mu} C_{\mu \mu'}
(\{\hat{F}_{\mu''}\},\epsilon)   - \hat{e}^c_{\mu'} = 0 
\end{equation}
Eq. (\ref{stationary}) determines the quasi stationary point
$\hat{F}_\mu(\epsilon)$ 
such,
that matrix $(C_{\mu \mu'}(\{\hat{F}_{\mu''},\epsilon))$ has an
eigenvalue $\hat{E}_0(\epsilon) = 1$ for $\epsilon<0$. 
$\epsilon <0$ indicates that  the
temperature is $T= T_c(1 - \epsilon) >T_c$.  
This stationary point $\hat{F}_\mu$ differs from $F^c_\mu$ in
order $\epsilon$! This can be proven by using an argument analogous
to deriving the $\beta$ - relaxation equation 
\cite{leshouches}. Defining $\hat{\delta}_\mu $ by $\hat{F}_\mu =
F^c_\mu  +
\hat{\delta}_\mu$ and using the property that $C_{\mu
  \mu'}(\{\hat{F}_{\mu''}\},\epsilon)$ is a smooth differentiable
function of $\epsilon$, Eq. (\ref{stationary}) can be written, in
leading order in $\hat{\delta}_\mu$ and $\epsilon$ as
\begin{equation} \label{stationary2}
0=  \sum_{\mu} \hat{e}^c_{\mu} (\sum_{\mu''} C_{\mu \mu' \mu''}
({\bf F}^c,0) \hat{\delta}_{\mu''} + \left. \frac{\partial
  C_{\mu \mu'} 
({\bf F}^c,\epsilon)}{\partial \epsilon} 
\right|_{\epsilon=0} \epsilon \;)   
\end{equation}
here we have in addition used that $\hat{e}^c_\mu$ is the left eigenvector of
the stability matrix at the critical point with eigenvalue
$E_0=1$. 
The tensor $C_{\mu \mu' \mu''}$ is the partial
derivative of the stability matrix with respect to $F_{\mu''}$. 
Eq. (\ref{stationary2}) immediately shows, that $\hat{\delta}_\mu$
is (at least) of order $\epsilon)$. Note, that the argument crucially
depends on the well known phenomenon, that the static quantities
like structure factors, which determine completely the stability
matrix $(C_{\mu \mu'}(\{F_{\mu''}\},\epsilon)$ 
are varying smoothly across the glass  transition. 

We now want to estimate the
number of iterations required to approach 
the quasi stationary point $\hat{\bf F}$. 
For that matter the
iterates $F_\mu^{(n)}$ are expanded around the quasi stationary point 
\[F_\mu^{(n)} = \hat{F}_\mu + \delta^{(n)}_\mu \]
Neglecting terms of order $(\delta_\mu^{(n)})^3$ the iteration
(\ref{a1}) can be rewritten as 

\begin{equation}\label{itexpan}
\delta^{(n+1)}_\mu = \Delta \hat{C}_\mu  + \hat{C}_{\mu \mu'}
\delta^{(n)}_{\mu'} + \frac{1}{2} C_{\mu \mu'
\mu''} \delta^{(n)}_{\mu'} \delta^{(n)}_{\mu''}
\end{equation}
where the quantities with $\hat{ }$ are taken at $
(\{\hat{F}_{\mu''}\},\epsilon)$. The slowing down of the iteration
is due to the component of $\delta_\mu$
along the critical direction $e^c$. This component can be extracted by
writing $\delta^{(n)} = a^{(n)} e^c$ and multiplying Eq. (\ref{itexpan})
with the left eigenvector $\hat{e}^c$ of the stability matrix $C$. The
resulting equation for $a^{(n)}$ is then 
\begin{equation} \label{difference}
a^{(n+1)} -a^{(n)} = \sigma + (\lambda -1) (a^{(n)})^2
\end{equation}
Here we made use of (\ref{stationary}) and  have normalized $e^c,
\hat{e}^c$ such that $\sum_{\mu}\hat{e}^c_\mu e^c_\mu=1$. The
quantities    $\sigma =
\sum_\mu \hat{e}^c_\mu \Delta C_\mu$ and $\lambda = 1 + 
\sum_\mu \hat{e}^c_\mu  C_{\mu \mu'
\mu''} \hat{e}^c_{\mu'} e^c_{\mu''}$ can be identified as the
separation parameter 
and the exponent parameter $0 <\lambda<1$ of the $\beta$ - relaxation theory,
respectively \cite{leshouches,foot4}. 
Since consecutive iterates are very close to each other in the
vicinity of $\hat{\bf F}$ we can
rewrite (\ref{difference}) as a differential equation. 
\begin{equation} \label{dgl}
-\dot{a} = (|\sigma| + ( 1 -\lambda) a^2)
\end{equation}

Lets assume we start the iteration at $t=0$ with $a(0)=a_0$. Then  the
solution of (\ref{dgl}) is 
\begin{equation} \label{dglsol}
a(t) = -\frac{\sqrt{|\sigma|}}{\sqrt{1-\lambda}}
\tan(\sqrt{(1-\lambda) |\sigma|} \; \;t -\arctan(a_0
\frac{\sqrt{1-\lambda}}{\sqrt{|\sigma|}}))   
\end{equation}

Since the iteration is initialized with a value for $\bf F$ slightly in
the glass, the initial deviation $a_0$ from $\hat{\bf F}$ is of order
$\sqrt{|\sigma|} = O(\sqrt{|\epsilon|})$. 
From Eq. (\ref{dglsol}) then
follows, that  we need a
number of iteration of the order $1/\sqrt{|\epsilon|}$ to approach
$\hat{\bf F}$ (i.e. $a(t) =0$), and will stay close to $\hat{\bf F}$
(i.e. $a(t) \le O(\sqrt{|\sigma|}$)
for the same amount of  iterations
before the iterates 
decay to zero. This proves our statement that the quasi stationary
solution of the iteration for $T>T_c$ is approached with the same
amount of iterations as the stable solution slightly below $T_c$. But
there the glass solution agrees with  ${\bf F}^c$ only up to order
$\sqrt{\epsilon}$ the quasistationary solution
$\hat{\bf F}$ however agrees with ${\bf F}^c$ up to order $|\epsilon| \ll
\sqrt{|\epsilon|}$.

\end{appendix}
\vspace{1cm}

\noindent
\underline{ Acknowledgment}: It is a pleasure to thank W. G\"otze for
a critical reading of the manuscript. We are grateful to
the Sonderforschungsbereich 262 for financial support.

\newpage

\bigskip

\bigskip 
 
\bigskip 

\bigskip 

\newpage

\bigskip

\bigskip 

\bigskip


\begin{thebibliography}{99}
\bibitem{foot1}Here we always restrict ourselves to the idealized mode
coupling theory, which neglects all ergodicity restoring processes
like e.g. hopping.
\bibitem{foot2}This notation is used to distinguish the molecular approach
from the site - site description.
\bibitem{foot4} Note that our definition of
$C_{\mu \mu' \mu''}$ differs from the one used in
\cite{leshouches} analogous to \cite{foot3}.
\end{thebibliography}
\end{document}